\documentclass[lettersize,journal]{IEEEtran}

\IEEEoverridecommandlockouts                              % This command is only needed if 
\usepackage{empheq}
\usepackage{amsfonts}
\usepackage{mathrsfs}  
\usepackage{times}
\usepackage{tikz}
\usepackage{bm}
\usepackage{graphicx}
\usepackage{multirow}
\usepackage{hyperref}
\usepackage{booktabs, makecell, tabularx, tabulary}

\usepackage{dsfont}
\usepackage{epstopdf}
\usepackage{caption}
 \usepackage{dsfont}
\usepackage{subcaption}
\usepackage{algorithm}
\usepackage{algpseudocode}
\usepackage{cite}
\newcommand{\zak}[1]{ ({\color{cyan}Zak: #1})}
\newtheorem{definition}{Definition}

\newtheorem{remark}{Remark}
\newtheorem{proposition}{Proposition}

\newtheorem{theorem}{Theorem}
\title{\LARGE \bf
 Neural Port-Hamiltonian Models for Nonlinear Distributed Control: An Unconstrained Parametrization Approach}

\author{Muhammad Zakwan and Giancarlo Ferrari-Trecate% <-this % stops a space
\thanks{This work was supported as a part of NCCR Automation, a National Centre of Competence in Research, funded by the Swiss National Science Foundation (grant number 51NF40\_225155) and the NECON project (grant number 200021-219431)}% <-this % stops a space
\thanks{Authors are with the Institute of Mechanical Engineering, Ecole Polytechnique Fédérale de Lausanne (EPFL), CH-1015 Lausanne, Switzerland {\tt\small \{muhammad.zakwan\}@epfl.ch }}%
}

\begin{document}

\maketitle
\thispagestyle{empty}
\pagestyle{empty}

% %%%%%%%%%%%%%%%%%%%%%%%%%%%%%%%%%%%%%%%%%%%%%%%%%%%%%%%%%%%%%%%%%%%%%%%%%%%%%%%%
\begin{abstract}
The control of large-scale cyber-physical systems requires optimal distributed policies relying solely on limited communication with neighboring agents. 
However, computing stabilizing controllers for nonlinear systems while optimizing complex costs 
remains a significant challenge.
Neural Networks (NNs), known for their expressivity, can be leveraged to parametrize control policies that yield good performance. However, NNs' sensitivity to small input changes poses a risk of destabilizing the closed-loop system. 
Many existing approaches enforce constraints on the controllers' parameter space to guarantee closed-loop stability, leading to computationally expensive optimization procedures.
To address these problems, we leverage the framework of port-Hamiltonian systems to design  continuous-time distributed control policies for nonlinear systems that guarantee closed-loop stability and finite 
$\mathcal{L}_2$ or incremental $\mathcal{L}_2$ gains, independent of the optimzation parameters of the controllers.
This eliminates the need to constrain parameters during optimization, allowing the use of standard techniques such as gradient-based methods. 
Additionally, we discuss discretization schemes that preserve the dissipation properties of these controllers for implementation on embedded systems. 
The effectiveness of the proposed distributed controllers is demonstrated through consensus control of non-holonomic mobile robots subject to collision avoidance and averaged voltage regulation with weighted power sharing in DC microgrids.

% \zak{Main points to discuss: 

% 3) admissible inputs for the discretization

% }
\end{abstract}

% \zak{
% \begin{enumerate}
%     \item Applications and comparison with Rens on theoretical and practical differences? Which one is more conservative? what is the benefit of nonlinear storage function? Empirical example? 
%     \item Can we say something for UAPs ? 
%     \item Add the reference of IFAC system identification talk with delta ISS and also criticize it 
%     \item Application to levels sets and approximations? Distributed examples?  
%     \end{enumerate}

% Michele Tucci 

% \begin{IEEEkeywords}
% Classification, Contraction theory, Hamiltonian systems, Machine learning,  Neural ODEs
% \end{IEEEkeywords}
%%%%%%%%%%%%%%%%%%%%%%%%%%%%%%%%%%%%%%%%%%%%%%%%%%%%%%%%%%%%%%%%%%%%%%%%%%%%%%%%
\section{Introduction}

Distributed control of \emph{large-scale} systems presents significant challenges even in seemingly basic scenarios due to the constrained flow of information in real-time. 
Particularly, Witsenhausen’s counter-example \cite{Witsenhausen} demonstrates that even under apparently ideal conditions (i.e. linear dynamics, quadratic loss, and Gaussian noise), a nonlinear distributed control policy can outperform the best linear one.
The work \cite{lessard2011quadratic} has provided necessary and sufficient condition, namely, Quadratic Invariance (QI), under which a distributed optimal controller is linear and can be designed by solving a convex optimization problem. 
However, real-world systems often violate the QI assumption due to inherent nonlinearities, non-convex control costs, or privacy limitations \cite{furieri2022distributed}. 
This necessitates venturing beyond linear control and exploring highly nonlinear distributed control policies, such as those parametrized by deep Neural Networks (NNs).

% \zak{Stability is an issue ??? }

NNs have proved their capabilities in learning-enabled control \cite{brunke2022safe, tsukamoto2021contraction, dawson2022safe, furieri2022distributed, Furieri2024learning,zakwan2024neuralcontraction}, and system identification \cite{nghiem2023physics, beintema2023deep, revay2023recurrent, zakwan2023physically, di2023simba, di2023stable} of non-linear dynamical systems.
Indeed, NN control has been applied in diverse contexts, such as robotics \cite{brunke2022safe}, epidemic models \cite{asikis2022neural}, safe path planning \cite{dawson2022safe}, and nonlinear consensus \cite{bottcher2022ai}.    
Existing approaches to NN control design also include
 modelling the system under control as a NN from data  \cite{hewing2019cautious, armenio2019model, bonassi2022recurrent, terzi2021learning}.
Nevertheless, NNs can be susceptible to small changes in their inputs~\cite{zakwan2022robust}. 
This fragility can readily lead to control policies jeopardizing closed-loop stability~\cite{revay2023recurrent}, thereby hindering their deployment in large-scale, safety-critical applications~\cite{dawson2022safe}.

In this paper, we leverage the well-established port-Hamiltonian (pH) system framework \cite{vanderSchaft2017} to provide an \emph{unconstrained parametrization} of distributed control policies that are inherently endowed with a finite $\mathcal{L}_2$ or an incremental $\mathcal{L}_2$ ($i\mathcal{L}_2$) gain. This allows casting  optimal control design into an unconstrained optimization problem that can be solved 
using standard gradient-based methods such as stochastic gradient descent or its variants. 
As a consequence, our approach eliminates the need for computationally expensive procedures, such as the projection of parameters or constrained optimization techniques, which are typically required to guarantee closed-loop stability~\cite{dawson2022safe}. 
Specifically, if both the underlying system to be controlled and the proposed controller satisfy the conditions of the small-gain theorem \cite{vanderSchaft2017}, our framework can ensure closed-loop stability both during and after the training.
Moreover, the learned distributed policies are optimal in the sense that they strive to minimize an arbitrary nonlinear cost function over a finite horizon.

{\bf Related work:} 
NNs have shown promise in designing both static and dynamic distributed control policies for large-scale systems.
Notably, Graph Neural Networks (GNNs) have achieved impressive performance in applications like vehicle flocking and formation flying \cite{yang2021communication, tolstaya2020learning, khan2020graph, gama2021graph} thanks to their inherent scalable structure.
However, guaranteeing stability with general GNNs remains challenging,  often requiring restrictive assumptions like linear, open-loop stable system dynamics or sufficiently small Lipschitz constants \cite{gama2021graph}.
Such limitations can be impractical, potentially leading to system failures during the training phase before an optimal policy can be found \cite{brunke2021safe,cheng2019end}. 
Some remedies to rectify this problem include improving an initial known safe policy iteratively, while imposing the constraint that the initial \emph{region of attraction} does not shrink \cite{berkenkamp2018safe,richards2018lyapunov,koller2018learning}, and 
 leveraging integral quadratic constraints to enforce the closed-loop stability \cite{pauli2021offset}. 
 However, these approaches explicitly constrain the optimization parameters of NNs, which may lead to infeasibility or hinder the closed-loop performance. 
 In contrast, our proposed method based on unconstrained parametrizations provides the same scalability as GNNs without imposing any constraints on the optimization parameters to guarantee closed-loop stability. 
 Previous works also explored stable-by-design control based on mechanical energy conservation \cite{abdulkhader2021learning, duong2021hamiltonian}, but these methods are limited to specific systems (e.g., with SE(3) dynamics). Our approach, instead, applies to a much wider range of nonlinear systems.

Recently, the notion of unconstrained parametrization has emerged for learning-enabled control, where a controller is parametrized such that it satisfy specific constraints (e.g., semi-definite constraints) \emph{by design}, i.e. without being constrained. This allows one to bypass  computationally expensive \emph{a posteriori} verification routines.
Based on this approach, the framework of Recurrent Equilibrium Networks (RENs) has been proposed in~\cite{revay2023recurrent}.
RENs are a class of discrete-time nonlinear dynamical models that enjoy built-in stability and robustness properties. Notably,  subsets of RENs can satisfy desired integral quadratic constraints regardless of their optimization parameters and ensure a finite $\mathcal{L}_2$ gain. 
Despite their flexibility, RENs face some key limitations. Firstly, they consider dynamics that are dissipative with respect to quadratic storage functions, potentially limiting their expressiveness for complex systems. Secondly, the unconstrained parametrization approach in \cite{revay2023recurrent, martinelli2023unconstrained}  cannot be directly applied to distributed systems where sparsity patterns in weight matrices are crucial. 
In contrast, our framework based on pH models allows the use of arbitrary nonlinear storage functions to capture more complex dynamics. Additionally, it seamlessly integrates desired sparsity patterns into the optimization parameters, enhancing flexibility without compromising stability. Since our framework is based on pH systems, a limitation of our controllers is that they have the same number of inputs and outputs. This is not required for RENs. 
Building on RENs, the work \cite{massai2023unconstrained} presents an unconstrained parametrization approach for interconnecting subsystems with finite $\mathcal{L}_2$ gain, while guaranteeing the $\mathcal{L}_2$ stability of the overall system. However, this approach is limited to quadratic storage functions for subsystems, constraining flexibility and generalization.
The work \cite{furieri2022distributed} presented a distributed framework based on pH systems that ensure passivity by design. However, it assumes that subsystems are in the pH form as well. Moreover, it does not guarantee finite $\mathcal{L}_2$ or  $i\mathcal{L}_2$ gains for the closed-loop system which is instead our main result.
Unlike passivity, a finite $\mathcal{L}_2$ gain guarantees stability even in the presence of external disturbances or modeling errors, which is crucial for safe operation in uncertain environments \cite{vanderSchaft2017, khong2018converse}.

{\bf Contributions:}
The main contributions of this paper can be summarized as follows: 
\begin{enumerate}
    \item We provide an unconstrained parametrization of a class of distributed controllers in the pH form that can seamlessly incorporate sparsity in their weight matrices and are inherently endowed with a finite $\mathcal{L}_2$ or  $i\mathcal{L}_2$ gain. 
     \item Our approach overcomes the limitation of being restricted to specific storage functions (e.g., quadratic), enabling its application to a broader range of nonlinear control problems.
     \item We illustrate how discrete gradient methods \cite{ehrhardt2018geometric} can be leveraged to preserve the dissipative properties when continuous-time pH controllers are discretized in time for implementation purposes.  
    \item We demonstrate the effectiveness of our controllers on a benchmark consensus problem for non-holonomic agents subject to a collision avoidance constraint, and weighted power sharing and average voltage regulation in DC microgrids.
\end{enumerate}

{\bf Organization:} 
Section \ref{Chap2:sec:prelims} presents the problem formulation. We then introduce unconstrained parametrizations of distributed controllers using pH models that guarantee a finite $\mathcal{L}_2$ gain (Section \ref{Chap2:sec:main_results}), or a finite  $i\mathcal{L}_2$ gain  (Section \ref{sec:incremental_l2}). Moreover,  Section \ref{Chap2:discretization_controllers} discusses the discretization methods to preserve these gains for practical implementation. Finally, the performance evaluation of our proposed pH controllers is presented in Section \ref{sec:experiments}, while Section \ref{Chap2:sec:conclusion} concludes the paper.

A preliminary version of this work appeared in~\cite{zakwan2024neural}, which is, however, substantially different from the present paper. Indeed, the following results were not provided:  (i) the unconstrained parametrization of a finite  $i\mathcal{L}_2$ gain (ii) the use of dissipation-preserving discretization schemes for implementation purposes, and (iii) a detailed proof of the unconstrained parametrization of pH models with finite $\mathcal{L}_2$ gains.

\begin{figure}
    \centering
    \includegraphics[width = 0.7\linewidth]{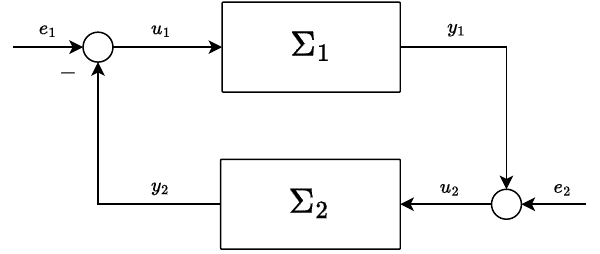}
    \caption{Standard feedback interconnection $\Sigma_1 \Vert_f \Sigma_2$.}
    \label{Chap2:fig:small_gain_theorem}
\end{figure}

{\bf General notation:} 
The set of non-negative real numbers is $\mathbb{R}_+$
and the standard Euclidean $2$-norm is denoted by $\Vert \cdot \Vert$. The identity matrix of size $n$ is denoted by $I_n$, a zero/null matrix of dimension $n \times m$ is given by $0_{n \times m}$, and $\mathds{1}$ is a vector of all ones with an appropriate dimension.
The maximal eigenvalue of a matrix ${A}$ is represented by $\bar{\lambda}({A})$. 
We represent the set of $\mathbb{R}^n$-valued Lebesgue square-integrable functions by $\mathcal{L}_2^n := \{ v : [0, \infty ) \rightarrow \mathbb{R}^n \vert \Vert v \Vert_2^2 := \int_0^\infty v(t)^\top v(t) dt < \infty\}$. We omit the dimension $n$ whenever it is clear from the context.
Then, for any two $v,w \in \mathcal{L}_2^n$, we denote the $\mathcal{L}_2^n$-inner product as $\langle v,w \rangle:= \int_0^\infty v(t)^\top w(t) dt$. Define the truncation operator $(P_\mathcal{T} v)(t) := v(t)$ for $t \leq \mathcal{T}$; $(P_\mathcal{T} v)(t) := 0$ for $t > \mathcal{T}$, and the extended function space $\mathcal{L}_{2e}^n := \{ v : [0, \infty ) \rightarrow \mathbb{R}^n \vert P_\mathcal{T} v \in \mathcal{L}_2, \forall \mathcal{T} \in [0,\infty)\}$. For any linear space $\mathcal{U}$ endowed with a norm $\Vert \cdot \Vert_{\mathcal{U}}$, we define a Banach space $\mathcal{L}_{2e}(\mathcal{U})$ that consists of all measurable functions $f:\mathbb{R}_+ \mapsto \mathcal{U}$ such that  $\int_0^\infty \Vert f(t) \Vert^2_{\mathcal{U}} dt < \infty$.
Throughout this paper, a system will be specified
by an input–output map $\Sigma: \mathcal{L}_{2e}^m \rightarrow \mathcal{L}_{2e}^p$ satisfying $\Sigma(0) = 0$. Given two systems $\Sigma_1$ and $\Sigma_2$, the standard negative feedback configuration between them is denoted by $\Sigma_1 \Vert_f \Sigma_2$, see Fig. \ref{Chap2:fig:small_gain_theorem}.
Let $\mathcal{G} = (\mathcal{V}, \mathcal{E})$ be an undirected graph with nodes $\mathcal{V} = \{ 1, \dots,N \}$ and edges $\mathcal{E}$, and let $\mathcal{P} \in \{0, 1\}^{N \times N}$ be the corresponding adjacency matrix.  
For a binary mask $\mathcal{M} \in \{0,1\}^{m \times n}$, we denote $\bm{W} \in \texttt{blkSparse}(\mathcal{M})$ if  $\bm{W}$ is a block matrix composed by $m \times m$ blocks and $\mathcal{M}_{i,j} = 0 \Rightarrow \bm{W}_{i,j} = 0$. $\bm{A} = \texttt{blkdiag}(A_i)$ represents a block-diagonal matrix with matrices $A_0, A_1,\dots, A_i$  on the diagonal.

{\bf Preliminaries:}
Consider the following non-linear system
\begin{subequations}
  \begin{empheq}[left={\Sigma :\empheqlbrace\,}]{align}
     \dot{x}(t)&=f\left(x(t), u(t)\right), \\
    y(t) &= h(x(t)) \;,
  \end{empheq}
\end{subequations}
where $x \in \mathcal{X} \subseteq \mathbb{R}^{n}$ is the state, $u \in \mathcal{U} \subseteq \mathbb{R}^{m}$ is the input, and $y \in \mathcal{Y} \subseteq \mathbb{R}^{p}$ is the output of the system $\Sigma$. Moreover, assume there exists a unique solution $x(t)$ on the infinite time interval $[0, \infty)$ of  \eqref{Chap2:eq:system1} for all initial conditions $x(0) \in \mathcal{X}$, and $u(\cdot) \in \mathcal{L}_{2e}(\mathcal{U})$. We recall some important results and definitions that are used to obtain the main results of this paper. 

\begin{definition}[Dissipativity, \cite{vanderSchaft2017}] \label{Chap2:def:dissipativity}The system $\Sigma$ is called dissipative w.r.t. to a supply rate $s : \mathcal{U} \times \mathcal{Y} \mapsto \mathbb{R}$, if there exists a smooth storage function $V:\mathcal{X} \mapsto \mathbb{R}_+$\footnote{For $\mathcal{X} = \mathbb{R}^{n}$, the storage function has to be radially unbounded, that is $V(x) \rightarrow \infty$, whenever $\Vert x \Vert \rightarrow \infty$ \cite[Theorem 3.2.4]{vanderSchaft2017}. } such that
$$
\dot{V}({x}(t)) \leq s(u(t),y(t)), \quad  \forall t \in \mathbb{R}_+ \;,
$$ or equivalently,
$$ V({x}(\tau)) - V({x}(0)) \leq \int_0^\tau s_i(u(t),y(t)) dt \;,
$$
for every input signal ${u}(t) \in \mathcal{U}$, output signal $y(t) \in \mathcal{Y}$ and $\tau \geq 0$. Moreover, the choice of the supply rate leads to different notions of dissipativity, for instance, 
\begin{enumerate}
    \item[$\bullet$] if $p = m$, and $s(u(t),y(t)) = u(t)^\top y(t)$, then system $\Sigma$ is passive; 
    \item[$\bullet$]  if $p = m$, and $s(u(t),y(t)) = u(t)^\top y(t) - \epsilon \Vert y(t)\Vert$, then system $\Sigma$ is $\epsilon$-output strictly passive for $\epsilon>0$;
    \item[$\bullet$]for some  non-negative constants $\gamma, b$,   if $s(u(t),y(t)) = \frac{\gamma^2}{2} \Vert u(t) \Vert - \frac{1}{2}\Vert y(t) \Vert$, then system $\Sigma$ has  a finite $\mathcal{L}_2$ gain, i.e. $\Vert y(t) \Vert \leq \gamma \Vert u(t) \Vert + b$.   
\end{enumerate}
% Note that the storage function ${V}(\cdot)$ can be interpreted as the stored ``energy" in the system w.r.t. a single point of neutral storage (minimum energy).

% Here,  the bold-faced signals $\bm{x}(t), \bm{u}(t)$, and $\bm{y}(t)$ are the stacked vectors defined as follows 
% \begin{equation}
%     \bm{x} = \begin{bmatrix}
%         x_i \\ \breve{x}_i
%     \end{bmatrix}, \bm{u} = \begin{bmatrix}
%         u_1 \\ \vdots \\ u_N 
%     \end{bmatrix}, \ \text{and} \ \bm{y} = \begin{bmatrix}
%         y_1 \\ \vdots \\ y_N 
%     \end{bmatrix} \;.
% \end{equation}  
\end{definition}

Consider the interconnection $\Sigma_1 \Vert_f \Sigma_2$ of two dissipative systems $\Sigma_1$, and $\Sigma_2$ in Fig. \ref{Chap2:fig:small_gain_theorem}. Then, one can leverage the following result to ensure closed-loop stability. 

\begin{theorem}[\cite{vanderSchaft2017}] \label{Chap2:thm:small_gain}
Consider the closed-loop system $\Sigma_1 \Vert_f \Sigma_2$ given in Fig. \ref{Chap2:fig:small_gain_theorem}.   
\begin{enumerate}
    \item[$\bullet$] (small gain condition) Assume the existence of  the $\mathcal{L}_2$ gains $\mathcal{L}_2(\Sigma_1) \leq \gamma_1$, and $\mathcal{L}_2(\Sigma_2) \leq \gamma_2$. Then, the closed-loop system $\Sigma_1 \Vert_f \Sigma_2$ is stable with an $\mathcal{L}_2$ gain $ \gamma_1 . \gamma_2$ if $\gamma_1 . \gamma_2 < 1$;
\item[$\bullet$]  (strict output passivity) Assume that, for any $e_1 \in \mathcal{L}_{2e}(\mathcal{U}_1)$ and $e_2 = 0$,  $\Sigma_1:\mathcal{L}_{2e}(\mathcal{U}_1) \rightarrow \mathcal{L}_{2e}(\mathcal{Y}_1)$ is $\epsilon_1$-output strictly
passive, and $\Sigma_2:\mathcal{L}_{2e}(\mathcal{U}_2) \rightarrow \mathcal{L}_{2e}(\mathcal{Y}_2)$ is passive. Then for $e_2 = 0$,  $\Sigma_1 \Vert_f \Sigma_2$  with input $e_1$ and output $y_1$ has an $\mathcal{L}_2$-gain $\leq 1 / \epsilon_1$.
\end{enumerate}
\end{theorem}

We recall that $\epsilon$-output strict passivity also implies a finite $\mathcal{L}_2$ gain not larger than $1 / \epsilon$ \cite{vanderSchaft2017}. 

% \zak{Moreover, the asymptotic stability of an equilibrium point $\bm{z}^\star := [\bm{x}^\star, \bm{\xi}^\star]$ of the closed-loop follows from the direct application of \cite[Proposition 3.2.12]{vanderSchaft2017}.}

While Theorem \ref{Chap2:thm:small_gain} guarantees  $\mathcal{L}_2$ stability of the closed-loop system, in several cases a stronger notion of stability, such as a finite  $i\mathcal{L}_2$ gain, is required \cite{vanderSchaft2017}.
Incremental stability \cite{angeli2002lyapunov} has garnered increasing interest in recent years due to its potential applications in the synchronization of chaotic systems \cite{stan2007analysis, tran2016incremental} and nonlinear circuits \cite{sepulchre2022incremental}. Moreover, incremental $\mathcal{L}_2$ stability implies that any two trajectories must eventually asymptotically converge to each other, regardless of their initial conditions \cite{arcak2016networks}.
Since incremental stability is an incremental dissipativity property, we formally recall the latter notion.

\begin{definition}[ Incremental Dissipativity, \cite{verhoek2023convex}] \label{Chap2:def:incremental_l2_gain}
    
    The system $\Sigma$ is called incrementally dissipative w.r.t. to a supply rate $s_{\Delta} : \mathcal{U} \times \mathcal{U} \times  \mathcal{Y} \times \mathcal{Y}\mapsto \mathbb{R}$, if there exists a smooth storage function $V_{\Delta}:\mathcal{X} \times \mathcal{X}  \mapsto \mathbb{R}_+$ such that  for any two trajectories $({x},{u},{y}), (\tilde{{x}},\tilde{{u}},\tilde{{y}})$
$$
\dot{V}_{\Delta}({x}(t), \tilde{x}(t)) \leq s_{\Delta}(u(t), \tilde{u}(t), y(t), \tilde{y}(t)), \ \forall t \in \mathbb{R}_+ \;,
$$ or equivalently,
\begin{align*}
  V_{\Delta}({x}(\tau), \tilde{x}(\tau)) &- V_{\Delta}({x}(0), \tilde{x}(0)) \\
&\leq \int_0^\tau s_{\Delta}(u(t),\tilde{u}(t), y(t), \tilde{y}(t)) dt \;,  
\end{align*}
for every pair of input signals ${u}(t), \tilde{u}(t) \in \mathcal{U}$, output signals $y(t), \tilde{y}(t) \in \mathcal{Y}$ and every $\tau \geq 0$. Moreover, the choice of supply rate leads to different notions of dissipativity, for instance, 
\begin{enumerate}
\item[$\bullet$]  if $p = m$, and $s_{\Delta}(u(t), \tilde{u}(t), y(t), \tilde{y}(t)) = (u(t) - \tilde{u}(t))^\top (y(t) - \tilde{y}(t))$, then system $\Sigma_{s}$ is incrementally passive;
    \item[$\bullet$]  if $p = m$, and $s_{\Delta}(u(t), \tilde{u}(t), y(t), \tilde{y}(t)) = (u(t) - \tilde{u}(t))^\top (y(t) - \tilde{y}(t)) - \epsilon_\Delta \Vert y(t) - \tilde{y}(t)\Vert$, then system $\Sigma$ is $\epsilon_\Delta$-output strictly incrementally passive for $\epsilon_\Delta>0$;
    \item[$\bullet$]  if $s_{\Delta}(u(t), \tilde{u}(t), y(t), \tilde{y}(t)) = \gamma^2_\Delta \Vert u(t) - \tilde{u}(t) \Vert - \Vert y(t) - \tilde{y}(t) \Vert$, then system $\Sigma$ has a finite  $i\mathcal{L}_2$ gain, i.e. $\Vert y(t) - \tilde{y}(t)\Vert \leq \gamma_\Delta \Vert u(t) - \tilde{u}(t) \Vert $ for a non-negative constant $\gamma_\Delta$, which is called the incremental gain.   
\end{enumerate}

\end{definition}
Note that $\epsilon_\Delta$-output strictly incremental passivity implies an $i\mathcal{L}_2$ gain of ${1}/{\epsilon_\Delta}$ \cite[{Proposition 2.2.22}]{vanderSchaft2017}.
The following result about closed-loop properties parallels Theorem \ref{Chap2:thm:small_gain}.

\begin{theorem}[\cite{vanderSchaft2017}]
\label{Chap2:thm2:inc_stab}
    Consider the closed-loop system $\Sigma_1 \Vert_f \Sigma_2$ given in Fig. \ref{Chap2:fig:small_gain_theorem}.   
\begin{enumerate}
    \item[$\bullet$] (incremental form of small gain condition) Assume the existence of the $i\mathcal{L}_2$ gains $\mathcal{L}_{2}^\Delta (\Sigma_1) \leq \gamma_{\Delta, 1}$, and $\mathcal{L}_2^\Delta(\Sigma_2) \leq \gamma_{\Delta,2}$. Then, the closed-loop system $\Sigma_1 \Vert_f \Sigma_2$ is stable with an $i\mathcal{L}_2$ gain $ \gamma_{\Delta,1} . \gamma_{\Delta, 2} < 1;$
\item[$\bullet$]  (strict output incremental passivity) Assume that, for any $e_1 \in \mathcal{L}_{2e}(\mathcal{U}_1)$ and $e_2 = 0$,  $\Sigma_1:\mathcal{L}_{2e}(\mathcal{U}_1) \rightarrow \mathcal{L}_{2e}(\mathcal{Y}_1)$ is $\epsilon_{\Delta, 1}$-output strictly
passive, and $\Sigma_2:\mathcal{L}_{2e}(\mathcal{U}_2) \rightarrow \mathcal{L}_{2e}(\mathcal{Y}_2)$ is incrementally passive. Then,  $\Sigma_1 \Vert_f \Sigma_2$ for $e_2 = 0$ with input $e_1$ and output $y_1$ has an $i\mathcal{L}_2$-gain $\leq 1 / \epsilon_{\Delta, 1}$.

\end{enumerate}
\end{theorem}

The conditions provided in Definition \ref{Chap2:def:incremental_l2_gain} for ensuring a finite $i\mathcal{L}_2$ gain lead to Hamilton-Jacobi inequalities, which are nonlinear infinite-dimensional partial differential constraints that are difficult to satisfy \cite{vanderSchaft2017}. Another approach, used in this paper, is to analyze the infinitesimal variation in the trajectories, as done in contraction theory \cite{verhoek2023convex}. To this end, we introduce the notion of differential dissipativity.
 \begin{definition}[Differential Dissipativity, \cite{verhoek2023convex}]\label{def:differential_dissipativity}

Consider the system $\Sigma$ and its variational dynamics 
\begin{subequations}
  \begin{empheq}[left={\Sigma_{\delta }:\empheqlbrace\,}]{align}
     \delta \dot{x}&= A(x,\delta x) \delta x + B(x, \delta x) \delta u , \nonumber \\
    \delta y &= C(x, \delta x) \delta x \;, \nonumber 
  \end{empheq}
\end{subequations}
 where $A(x,\delta x):= \frac{\partial }{\partial x} f\left(x, u\right)$, $B(x,\delta x) := \frac{\partial }{\partial u} f\left(x, u\right)$, $C(x, \delta x) := \frac{\partial }{\partial x} h\left(x\right)$ are matrix-valued functions. Then, $\Sigma$ and $\Sigma_{\delta}$
are called differentially dissipative w.r.t.   to the supply function ${s}_{\delta} : \mathcal{U} \times \mathcal{Y} \rightarrow  \mathbb{R}$, if there exists a smooth storage function $V_{\delta} : \mathcal{X} \times \mathcal{X} \rightarrow \mathbb{R}^+$, with $V_{\delta}(\cdot,0) = 0$ such that 
 \begin{align*}
     V_{\delta}({x}(t_1), \delta {{x}}(t_1)) &- V_{\delta} ({x}(t_0), \delta {{x}}(t_0)) \\
     &\leq \int_{t_0}^{t_1} s_{\delta} (\delta {u}(t), \delta {y}(t)) d t
 \end{align*}
 for all $t_0, t_1 \in \mathbb{R}$ with $t_0 \leq t_1$ and for all $(\delta {x}, \delta {u})$.
\end{definition}

\section{Problem formulation} \label{Chap2:sec:prelims}

In this section, we formally formulate the distributed control problem of interest. To this end, we consider a network $\Sigma_s$ of $N$ coupled nonlinear subsystems, each endowed with a feedback control policy. Let $\mathcal{G}_s = (\mathcal{V}_s, \mathcal{E}_s)$ represent the undirected graph associated with the couplings among subsystems, and let $\mathcal{P}_s$ be its corresponding adjacency matrix. We assume each subsystem is governed by
\begin{subequations}
  \begin{empheq}[left={\Sigma_{s,i}:\empheqlbrace\,}]{align}
     \dot{x}_i(t)&=f_i\left(x_i(t), \breve{x}_i(t), u_i(t)\right), \label{Chap2:eq:system1}\\
    y_i(t) &= h_i(x_i(t)), \qquad \quad   \ \forall i \in \mathcal{V}_s \;, \label{Chap2:eq:system2}
  \end{empheq}
\end{subequations}
where $x_i \in \mathcal{X}_i \subseteq \mathbb{R}^{n_i}$ is the state, $u_i \in \mathcal{U}_i \subseteq \mathbb{R}^{m_i}$ is the input, and $y_i \in \mathcal{Y}_i \subseteq \mathbb{R}^{p_i}$ is the output of the subsystem $\Sigma_{s,i}$, respectively.\footnote{The sets $\mathcal{X}_i$, $\mathcal{U}_i$, and $\mathcal{Y}_i$ are assumed to be non-empty.} 
% $\mathcal{U}_i$ and $\mathcal{Y}_i$ are linear spaces of dimension $m_i$, respectively $p_i$. 
We define $\breve{x}_i$ as a stacked vector of states of the $1$-hop neighbors of subsystem $i$ according to $\mathcal{G}_s$, i.e. all subsystems that influence $x_i$.
Moreover, we assume there exists a unique solution trajectory on the infinite time interval $[0, \infty)$ of  \eqref{Chap2:eq:system1} for all initial conditions $x_i(0) \in \mathcal{X}_i$, and $u_i(\cdot) \in \mathcal{L}_{2e}(\mathcal{U}_i)$.
% When the number of subsystems $N$ is very large, we say that network system $\Sigma_s$ is large-scale.
% One can write the system \eqref{eq:system} in a compact form as 
% \begin{equation} \label{Chap2:eq:system_full}
%     \dot{\bm{x}}(t) = \bm{F}(\bm{x}(t), \bm{u}(t)), \quad \bm{y}(t) = \bm{h}(\bm{x}(t))
% \end{equation}
% where the bold faced signals $\bm{x}(t), \bm{u}(t)$, and $\bm{y}(t)$ are the stacked vectors as follows
% \begin{equation}
%     \bm{x} = \begin{bmatrix}
%         x_i \\ \breve{x}_i
%     \end{bmatrix}, \bm{u} = \begin{bmatrix}
%         u_1 \\ \vdots \\ u_N 
%     \end{bmatrix}, \ \text{and} \ \bm{y} = \begin{bmatrix}
%         y_1 \\ \vdots \\ y_N 
%     \end{bmatrix} \;, 
% \end{equation}
% respectively. The matrix-valued functions in \eqref{eq:system_full} are defined as $\bm{F} = [f_1, \dots, f_N]^\top$, and $\bm{h} = blkdiag(h_1, \dots, h_N)$.

The distributed control of large-scale systems presents a major challenge: local controllers at each subsystem $u_i(t)$ can only access real-time information from a limited set of neighbors, dictated by a communication network, which is modeled as an undirected graph $\mathcal{G}_c = (\mathcal{V}_s, \mathcal{E}_s)$. This network is equivalently represented by the adjacency matrix $\mathcal{P}_c \in \{0, 1\}^{N \times N}$, where $\mathcal{P}_{c_{i,i}} = 1$ for every $i \in \mathcal{V}_s$.
% \footnote{In our setting, we assume equal number of sub-systems and sub-controllers, i.e., the cardinality of $\#\mathcal{V}_s = \# \mathcal{V}_c$.}
% The main challenge of the distributed control for large-scale systems like $\Sigma_s$ is that the local control inputs $u_i(t)$ can only receive real-time information from a limited subset of neighboring subsystems according to a communication topology described by a graph $\mathcal{G}_c$ and the accompanying adjacency matrix $\mathcal{P}_c \in \{0, 1\}^{N \times N}$ such that $\mathcal{P}_{c_{i,i}} = 1$ for every $i \in \mathcal{V}$. 
\begin{figure}
    \centering
    \includegraphics[scale = 0.5]{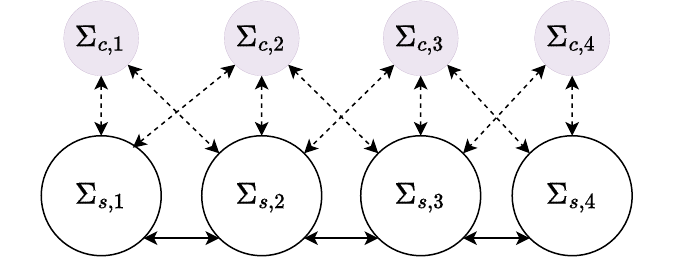}
    \caption{An example of an interconnected system $\Sigma_s$ and a distributed controller $\Sigma_c$ for $N = 4$.
    The solid lines represent interactions between the subsystems of $\Sigma_s$, and the dashed lines represent the flow of information between the system $\Sigma_s$ and the controller $\Sigma_c$.}
    \label{Chap2:fig:distributed}
\end{figure}
 In this paper, our goal is to develop distributed dynamic feedback controller $\Sigma_c$ represented by the pairs $(\chi_i(\cdot),\pi_i(\cdot)), i \in \mathcal{V}_s$ defining local controllers with some parameters $\theta_i \in \mathbb{R}^{d_i}$
\begin{equation} \label{Chap2:eq:cont_org}
\Sigma_{c,i}: \begin{cases}
        \begin{aligned}
        \dot{\xi}_i(t) &= \chi_i(\xi_i(t), \breve{y}_i(t), \theta_i), \quad \xi_i \in {\Xi}_i \subseteq \mathbb{R}^{q_i},\\
        u_i(t) &= \pi_i(\xi_i(t),\breve{y}_i(t), \theta_i) \;,
    \end{aligned}
\end{cases}
\end{equation}
where $\xi_i$ is the state of $\Sigma_{c,i}$, and $\breve{y}_i(t)$ is a stacked vector of outputs from the neighboring subsystems based on the communication graph $\mathcal{G}_c$. 
An example of a communication graph between the distributed controller $\Sigma_c$ and the distributed system $\Sigma_s$ is shown in Fig. \ref{Chap2:fig:distributed}.

Particularly, in this paper, we are interested in controllers $\Sigma_{c,i}$ that are dissipative,  endowed with some finite $\mathcal{L}_2$ or $i\mathcal{L}_2$ gains, and admit an \emph{unconstrained parametrization}, i.e. for a given $\gamma (\gamma_\Delta)> 0$,  
\begin{equation}
\label{Chap2:unconstrained_ppt}
    \Sigma_{c,i} \ \text{has finite $\mathcal{L}_2$ ($i\mathcal{L}_2$) gain $\gamma \ (\gamma_\Delta)$ \emph{for all}} \ \theta_i \in \mathbb{R}^{d_i} \;.
\end{equation}
The term ``unconstrained" refers to the fact that prescribed bounds on the $\mathcal{L}_2 (i \mathcal{L}_2)$ gains, i.e.  the property \eqref{Chap2:unconstrained_ppt} must be  guaranteed irrespectively of the value of $\theta_i \in \mathbb{R}^{d_i}$. In fact, the main result of this paper is to use pH systems for defining sets of controllers verifying \eqref{Chap2:unconstrained_ppt}. 
We define $\bm{\theta} = (\theta_1, \dots, \theta_N)$.
Moreover, the control policies $\Sigma_c$ should be \emph{optimal} in the sense that they minimize an arbitrary  real-valued cost function 
\begin{equation} \label{Chap2:eq:loss_function}
    c(\bm{x}(t),\bm{u}(t)) = \frac{1}{T} \int_0^T \ell(\bm{x}(t), \bm{u}(t)) dt
\end{equation}
for a finite horizon $T \in \mathbb{R}_+$, where $c$ is differentiable almost everywhere.  The bold-faced signals $\bm{x}(t) = [x_1^\top, \dots, {x}_N^\top]^\top, \bm{u}(t) = [u_1^\top, \dots, {u}_N^\top]^\top$ represent concatenated local states and local inputs, respectively. 
% Hence, the ODC boils down to find optimal parameters $\bm{\theta} = (\theta_1, \dots, \theta_N)$ by solving the following optimization problem
% \begin{align}
%     \min_{\bm{\theta}} \quad &c(\bm{x},\bm{u},\bm{\theta}) \\
%     \text{s.t.} \quad &\text{system dynamics \eqref{eq:system}} \nonumber \\ 
%       &\dot{\xi}_i(t) = \chi_i(\xi_i, \breve{y}_i(t), \theta_i), \label{Chap2:eq:cont_1}\\
%       &u_i(t) = \pi_i(\xi_i,\breve{y}_i(t),\theta_i), \quad \forall i \in \mathcal{V}, \label{Chap2:eq:cont_2}
% \end{align}

Besides designing optimal control policies, ensuring the stability of the closed-loop system $ \Sigma_s \Vert_f \Sigma_c$ is equally crucial. 
% since
% the closed-loop $ \Sigma_s \Vert_f \Sigma_c$ can become unstable leading to system failure and safety concerns \cite{furieri2022distributed}.
To address this issue, in the remainder of this paper, we focus on achieving several notions of closed-loop stability, such as $\mathcal{L}_2$ or $i\mathcal{L}_2$ stability based on the small-gain theorem, or strictly output passivity (Theorem \ref{Chap2:thm:small_gain}, and \ref{Chap2:thm2:inc_stab}). 
In a nutshell, these conditions enforce a prescribed $\mathcal{L}_2$ or $i\mathcal{L}_2$ gain on the controller $\Sigma_c$ for guaranteeing closed-loop stability. 
% Furthermore, we are interested in the notion of asymptotic $\mathcal{L}_2$ stability of the closed-loop $ \Sigma_s \Vert_f \Sigma_c$, according to the small-gain theorem.
% Before providing the formal statement of the small-gain theorem, let us recall the definition of $\mathcal{L}_2$-gain for nonlinear systems of the form $\Sigma_c$, for the sake of completeness.
% \begin{definition} \label{Chap2:def:l2_gain}
%    The nonlinear system $\Sigma_c$ has a finite $\mathcal{L}_2$-gain $\mathcal{L}_2(\Sigma_c) \leq \gamma_c$ if it is dissipative with respect to the supply rate $s(\bm{u}, \bm{y})=\gamma^2_c\|\bm{u}\|^2-\|\bm{y}\|^2$; that is, there exists a storage function ${V}: \mathbb{R}^n \rightarrow \mathbb{R}^{+}$ such that 
% $$
% \dot{V}(\bm{x}) \leq s(\bm{u}(t), \bm{y}(t)), \quad  \forall t \in \mathbb{R}^+
% $$
% or, equivalently, for $t \in \mathbb{R}^+$ and $\bm{x}_0 \in \mathbb{R}^n$,
% $$
% {V}(\bm{x}(\tau))- {V}\left(\bm{x}(0)\right) \leq \int_0^\tau s( \bm{u}(t), \bm{y}(t)) dt \;,
% $$ 
% for some $\tau \geq 0$.
% \end{definition}
% add a subtract 
% \zak{How we solve optimal control prob}
% \zak{How the discretization works? and the training of the controller, 
% clean the code and upload}

% Note that Theorem 1 requires that both the system and the controller are endowed with finite $\mathcal{L}_2$ gains. However, there is more relaxed version, namely, Passivity theorem, which states that given the system $\Sigma_s$ is passive and the controller $\Sigma_c$ is strictly output passive, then the closed-loop has a finite $\mathcal{L}
% _2$ gain. See \cite[Theorem2.2.15]{vanderSchaft2017} for more details.  
% In conclusion, our objective is to train neural control policies such that i) they are constrained by the availability of information, ii) they are optimal in the sense of minimizing a generic nonlinear cost function, and iii) guarantee a finite $\mathcal{L}_2$ gain of the neural closed-loop $\Sigma_s \Vert_f \Sigma_c$. 

% Altogether, our goal is to train control policies for large-scale systems that address three key requirements:
% \begin{enumerate}
%     \item[i)] Limited information access: The policies must operate with restricted local information.
%     \item[ii)] Optimal performance: They achieve optimal behavior by empirically minimizing an arbitrary user-defined cost function.
%     \item[iii)] Guaranteed stability of the closed-loop system.
% \end{enumerate} 
The overall problem can be formulated as the following Optimal Control Problem (OCP)
\begin{align}
\label{Chap2:opt_prob}
     \min_{\bm{\theta}} \quad & \frac{1}{S} \sum_{k = 1}^{S}c(\bm{x},\bm{u};\bm{\theta}, \bm{x}_0^k, \bm{\xi}_0^k) \\
    \text{s.t.} \quad &\text{system dynamics $\Sigma_s$}  \nonumber \\ 
      &\text{controller dynamics $\Sigma_c$}  \nonumber \\ \ & \Sigma_c \ \text{has a prescribed finite} \ \mathcal{L}_2, \nonumber   \\
      &\text{or} \ i\mathcal{L}_2 \ \text{gain}, \forall \bm{\theta} \in \mathbb{R}^{\bm{d}} \label{Chap2:const:l2} \;,
\end{align}
where $\bm{x}^k_0, \bm{\xi}^k_0, k = 1, \dots, S$ are samples of initial conditions for $\Sigma_s$, and $\Sigma_c$, respectively.\footnote{Note that the empirical average in the cost is an approximation of the exact and often uncomputable average $\mathbb{E}_{\bm{x}_0 \in \Omega_s, \bm{\xi}_0 \in \Omega_c}[c(\bm{x}, \bm{u}; \bm{\theta}, \bm{x}_0, \bm{\xi}_0)]$ defined w.r.t. a probability distribution over the initial conditions.} 
% The primary challenge of this OCP lies in finding the parameters $\bm{\theta}$ such that the distributed controller $\Sigma_c$ has a finite $\mathcal{L}_2$, or incremental $\mathcal{L}_2$ gain without imposing that $\bm{\theta}$ is constrained in a subset of $\mathbb{R}^{\bm{d}}$. 
% Before diving into this problem, the following section presents Hamiltonian models of the  controllers that will be leveraged to guarantee closed-loop stability and will be used throughout the paper. 

\begin{remark}[Time-varying parameters]
\label{Chap2:remark:time_varying_weights}
In several applications of nonlinear OCP, time-varying optimization parameters provide additional degrees of freedom to minimize an arbitrary cost function during a transient time over a finite horizon $T$. For instance, a mechanical robot performs some complex maneuvers for a finite time and then asymptotically reaches an equilibrium point. Therefore, one can choose a time-varying parametrization as follows:
\begin{equation*}
\bm{\theta}(t) = 
    \begin{cases}
        \bm{\theta}(t) \ \text{for} \ t \in [0, T], \\
        \bm{\theta}(T) \ \text{for} \ t > T \;,
    \end{cases}
\end{equation*}
where the optimization parameters $\theta(T)$ are frozen for $t > T$ and are computed such that $\Sigma_c$ has a finite $\mathcal{L}_2$, or incremental $\mathcal{L}_2$ gain.  
We demonstrate the efficacy of using time-varying weights by showing the consensus of non-holonomic agents in Section \ref{Chap2:example1}. 
\end{remark}

% As mentioned before, the nonlinear optimal control problem can be solved via standard packages, usually used in Model Predictive Control (MPC), such as ALADIN, and ADMM. In this paper, we leverage well-established Pytorch package to solve the problem \eqref{Chap2:opt_prob}-\eqref{Chap2:const:l2} as NN training problem, where the optimization parameters $\bm{theta}$ are trainable parameters. We defer the reader to Appendix \ref{Chap2:sec:training} for a brief discussion on the training of these parameters.}

\subsection{Port-Hamiltonian controllers} \label{Chap2:controller_structure}

% To satisfy the property in \eqref{Chap2:const:l2}, two common approaches, as outlined in \cite{dawson2022safe}, involve either constrained optimization or the projection of $\bm{\theta}$ onto the set $\Theta_{\mathcal{L}2}$ (or $\Theta_{\Delta \mathcal{L}_2}$ for incremental gains). These approaches ensure that the controller $\Sigma_c$ exhibits the prescribed finite $\mathcal{L}_2$ or incremental $\mathcal{L}_2$ gain, respectively.
% However, these methods can be computationally burdensome, and limiting the class of controllers that can be used. 
% To circumvent this issue, 
% we propose an unconstrained parametrization approach that involves designing a class of input-output operators possessing a finite $\mathcal{L}_2$, or incremental $\mathcal{L}_2$ gain for any choice of parameters $\bm{\theta}$. 
% This enables one to seamlessly employ unconstrained optimization methods such as stochastic gradient descent and its variants to solve \eqref{Chap2:opt_prob}-\eqref{Chap2:const:l2}.  To achieve this goal,
% we leverage the well-established pH framework \cite{vanderSchaft2017} to provide unconstrained parametrizations of the distributed controllers $\Sigma_c$ with guaranteed finite $\mathcal{L}_2$, or incremental $\mathcal{L}_2$ gains. This enables one to seamlessly employ unconstrained optimization methods such as stochastic gradient descent and its variants to solve \eqref{Chap2:opt_prob}-\eqref{Chap2:const:l2}.

We consider local controllers \eqref{Chap2:eq:cont_org} in the form
\begin{equation}
\label{Chap2:eq.controller_individual}
\Sigma_{pH,i}: \begin{cases}
\begin{aligned}
      \dot{{\xi}_i}(t)& =\left[{J}_{i}- (\alpha {I} + {\Lambda}_i)\right] \frac{\partial H_{i}}{\partial {\xi_i}}({\xi_i}, {{\vartheta_i}})+ {G}_{i} \breve{y}_i(t) \\
    {u}_i(t) & = {G}_{i}^\top \frac{\partial H_{i}}{\partial {\xi_i}}({\xi_i},{\vartheta_i}),  \qquad \forall i \in \mathcal{V}_s \;,
\end{aligned}
\end{cases}
\end{equation}
where the interconnection matrix ${J}_{i}=-{J}_{i}^{\top}$ is skew-symmetric, and the damping is determined by a diagonal matrix $(\alpha I_{q_i}+ \Lambda_i) >0$, where $\alpha >$ is a constant, and $\Lambda_i := \operatorname{diag}(e^{r}) \in \mathbb{R}^{q_i}_+$ for some free parameter vector $r \in \mathbb{R}^{q_i}$. The continuously differentiable and radially unbounded Hamiltonian function $H_i: \mathbb{R}^{q_{i}} \rightarrow \mathbb{R}_+$ of  $\Sigma_{pH,i}$ represents the energy of  $i^{th}$ controller  and is endowed with some parameters $\vartheta_i$. Moreover, we collect all the optimization parameters of $i^{th}$ sub-controller in a set $\theta_i :=\{ J_i, \Lambda_i, \vartheta_i, G_i \}$.  
% Moreover, in \eqref{Chap2:eq.controller_individual}, the matrices $J_i, \Lambda_i$, and $G_i$ can also be made trainable. 
For the sake of presentation, the distributed controller \eqref{Chap2:eq.controller_individual} can be compactly written as   
\begin{equation}
\Sigma_{pH}: \begin{cases}
\begin{aligned}
      \dot{\bm{\xi}}(t)& =\left[\bm{J}_{c}- (\alpha \bm{I} + \bm{\Lambda})\right] \frac{\partial H_{c}}{\partial \bm{\xi}}(\bm{\xi}, {\bm{\vartheta}})+ \bm{G}_{c} \bm{y}(t) \\
    \bm{u}(t) & = \bm{G}_c^\top \frac{\partial H_{c}}{\partial \bm{\xi}}(\bm{\xi},\bm{\vartheta}) \;,
\end{aligned}\label{Chap2:eq.Controller}  
\end{cases}
\end{equation}
where $\bm{\xi} \in \Xi_1 \times \cdots \times \Xi_{N} \subseteq \mathbb{R}^{n_c}, \bm{u}\in \mathcal{U}_{c,i} \times \cdots \times \mathcal{U}_{c,N} \subseteq \mathbb{R}^m, \bm{y} \in \mathcal{Y}_{c,i} \times \cdots \times \mathcal{Y}_{c,N} \subseteq \mathbb{R}^m$ are stacked vectors of controller states, outputs, and inputs, respectively. The block matrix $\bm{J}_{c}= \texttt{blkdiag}(J_i)$ is skew-symmetric, and $(\alpha \bm{I} + \bm{\Lambda})$,\footnote{Note that the parametrization $\alpha \bm{I} + \bm{\Lambda}$ is redundant, as the same set of matrices can be obtained by using $\bm{\Lambda}$ only. However, it makes it easier to enforce specific dissipativity properties as shown later in Section \ref{Chap2:L2operators}. } where $\bm{\Lambda} = \texttt{blkdiag}(\Lambda_i)$ is diagonal by construction, and 
the input matrix $\bm{G}_c = \texttt{blkSparse}(\mathcal{P}_c)$ is  full rank.\footnote{
Decentralized control is achieved by setting $\bm{G}_c = \texttt{blkdiag}(G_i)$ in \eqref{Chap2:eq.Controller}, making each sub-controller independent of the state of other subsystems or sub-controllers.}
Moreover, the  Hamiltonian function $H_c: \mathbb{R}^{n_c} \rightarrow \mathbb{R}_+$ of  $\Sigma_{pH}$ is the algebraic sum of all $N$ sub-controllers' energies $H_i$, i.e. $H_c (\bm{\xi}, \bm{\vartheta}) = \sum_{i \in \mathcal{V}_s} H_i(\xi_i(t), \vartheta_i)$. Finally, we gather all the optimization parameters of the distributed controller \eqref{Chap2:eq.Controller} in a set $\bm{\theta} := \{\bm{J}, \bm{\Lambda}, \bm{\vartheta}, \bm{G}_c  \}$.
% Since the controller \eqref{Chap2:eq.Controller} is endowed with some trainable parameters and NNs we refer to \eqref{Chap2:eq.Controller} as a NN controller and this structure will be used throughout the paper.
\begin{remark}[Selection of Hamiltonian]\label{Chap2:remark_hamiltonian}
    We impose minimal restrictions on the Hamiltonian functions $H_i({\xi_i}, {\vartheta}_i)$, i.e., differentiability and radial unboundedness. This flexibility allows using diverse parametrizations including simple quadratic functions, and NNs (as in \cite{galimberti2023hamiltonian, zakwan2023universal} for representing $H_i$). Several NNs  satisfy the mild assumptions above, for example:
    \begin{itemize}
        \item [$\bullet$] Multi-Layered Perceptron (MLPs) with  $\operatorname{SmoothReLU}(\cdot)$ activation functions $\tilde{\sigma}(\cdot)$
        \begin{align*}
        H_i(\xi_i, \vartheta_i) &= z_{L} \\
    z_1 &= \tilde{\sigma}_{1} \left ( W_{1} \xi_i +b_{1} \right ) \nonumber \\ 
    z_\ell &= \tilde{\sigma}_{\ell} \left ( W_{\ell}z _{\ell -1}+b_{\ell} \right ), \ \ell = 2, \cdots, L \;,     
\end{align*}  
% \begin{equation} \label{eq:controller}
%     u_\theta(x) = W_N\sigma_{N} \left (\cdots \sigma _{2}\left ( W_{2}\sigma_{1} \left ( W_{1}x+b_{1} \right )+b_{2} \right )\cdots  \right ),
% \end{equation}
where $L \geq 1$ represents the depth of the NN, and the parameters are $\vartheta_i := \{W_L, \cdots, W_1, b_1,\cdots, b_L \}$.
        \item Input Convex NNs (ICNNs) \cite{amos2017input} \begin{equation}
        \label{Chap2:eq:icnns}
        H_i({\xi_i},{\vartheta_i})  := {\hat{\sigma}}(\varphi_L({\xi_i})-\varphi(0)) +   \epsilon {\xi}^\top_i {\xi}_i \;, 
    \end{equation}
    with $\varphi_L$  defined as 
    \begin{align*}
        {z}_1 &= \hat{\sigma}_1({W}_1 {\xi}_i + b_1) \\
        {z}_{\ell} &= \hat{\sigma}_\ell({U}_\ell {z}_{\ell-1} + {W}_\ell {\xi_i} + {b}_\ell), \ \ell = 2, \cdots, L \\ 
        \varphi_L({\xi}_i) &= {z}_L \;,
    \end{align*}
where $\hat{\sigma}(\cdot)$ is a convex,  monotonically non-decreasing, non-negative activation function, such as $\operatorname{SmoothReLU}(\cdot)$, ${W}_\ell$ are real-valued weights, ${U}_\ell$ are positive weights, ${b}_\ell$ are real-valued biases, and $\epsilon >0$ is a small positive constant. In this case, we define  $\vartheta_i := \{W_1, \dots, W_L, U_2, \dots, U_L, b_1, \dots, b_L \}$.
        \item Residual-like Networks \cite{he2016deep} \begin{equation}
        \label{Chap2:eq:icnns}
        H_i({\xi_i},{\vartheta_i})  :=  W_o z_L({\xi_i}) + \epsilon \xi_i^\top \xi_i\;, 
    \end{equation}
    with $z_L$ defined as 
\begin{align*}
        {z}_1 &= {\sigma}_1({W}_1 {\xi}_i + b_1) \\
        {z}_{\ell} &= {\sigma}_\ell({W}_\ell {z}_{\ell-1} +  {b}_\ell), \ \ell = 2, \cdots, L \;,
    \end{align*}
    where $\sigma(\cdot)$ is a differentiable positive activation function, such as $\operatorname{sigmoid(\cdot)}$, $\operatorname{SmoothReLU(\cdot)}$, and $\operatorname{Softplus}(\cdot)$, $\epsilon >0$ is a small positive constant, and $\vartheta_i := \{W_o, W_1, \dots, W_L, b_1, \dots, b_L  \}$.
\end{itemize}
Since all the aforementioned activation functions are differentiable and non-negative, the Hamiltonian $H_i$ will also be differentiable and positive definite. Moreover, the radial unboundedness of the Hamiltonian is ensured by the use of unbounded activation functions for MLPs and the inclusion of quadratic terms $\epsilon \xi^\top_i \xi_i$ in ICNNs and ResNet-like NNs.
\end{remark}

 In the sequel, we consider NN parametrizations of $H_i$ because of their expressivity. Importantly, our results hold irrespective of the specific NN choice.
\subsection{Solving the OCP \eqref{Chap2:opt_prob}-\eqref{Chap2:const:l2} }

 There are various optimization methods for solving the finite-horizon OCP \eqref{Chap2:opt_prob}-\eqref{Chap2:const:l2}, including Sequential Quadratic Programming (SQP) \cite{nocedal1999numerical}, the Alternating Direction Method of Multipliers (ADMM) \cite{wang2019global}, the Augmented Lagrangian Alternating Direction Inexact Newton (ALADIN) method \cite{houska2016augmented}, and nonlinear interior point methods such as IPOPT \cite{biegler2009large}. In this paper, we leverage the well-established BackPropagation-Through-Time (BPTT) approach \cite{werbos1990backpropagation, NeuralODEs} to solve \eqref{Chap2:opt_prob}-\eqref{Chap2:const:l2} by reducing it to a NN training problem, where the optimization parameters $\bm{\theta}$ are referred as trainable parameters. 
This approach offers several advantages, including improved scalability through GPU parallelization,  the ability to parametrize highly nonlinear functions via NNs, the potential to deal with high dimensionality in both parameter and state space, and to cater longer time horizons \cite{di2023simba}.
We defer the reader to Appendix \ref{Chap2:sec:training} for a brief summary of BPTT. However, we highlight that the unconstrained parametrization of controllers with a finite $\mathcal{L}_2$ or $i\mathcal{L}_2$ gain in the following sections is independent of the chosen optimization method.

% A gooddoption would be to collect in remark 1 the following material
% - say that H can be parametrized using very different classes of functions. in the sequel, however, you consider NN parametrizations in view of their expressivity (that can be motivated using universal approximation theorems. if they are available)
% - review some relevant classes of NN that can be used for parametrizing H

% \zak{A small paragraph on the training of the closed loop from a a NN ode perspective, the connection with the NOC and NN training??}

\section{$\mathcal{L}_2$-stable Hamiltonian controllers} \label{Chap2:sec:main_results}

%To circumvent this, we propose a free parametrization approach, which involves designing a class of input-output operators that inherently possess a finite L2 gain for any choice of weight matrices θ. This allows us to employ unconstrained optimization methods like stochastic gradient descent and its variants. To achieve this, we leverage the well-established port-Hamiltonian framework to parametrize controllers Σ with guaranteed finite L2 gains.

% \subsection{An unconstrained parametrization of $\mathcal{L}_2$ operators}

\label{Chap2:L2operators}
% \section{Main Results} \label{Chap2:sec:main_results}
% Our main result concerns the free parametrization of the NN controller Σ 
% pH. This ensures a finite 
%   gain regardless of the specific values assigned to its trainable parameters. These parameters, represented by θ, consist of the weight matrices . .
The following result presents a parametrization of the controller $\Sigma_{pH}$ that guarantees a finite $\mathcal{L}_2$ gain
 regardless of the choice of the parameters $\bm{\theta}$ in \eqref{Chap2:eq.Controller}. The proof is provided in Appendix \ref{Chap2:thm2_proof}.

% In the sequel, we provide our main result on the free parametrization of NN controller $\Sigma_{pH}$ such that it has finite $\mathcal{L}_2$ gain for all choices of trainable parameters $\bm{\theta}$, where $\bm{\theta}$ denotes the set of weight matrices $\{ \bm{J}_c, \bm{\Lambda}, \bm{G}_c, \theta \}$. The detailed proof of this result is provided in the Appendix.

\begin{theorem} \label{Chap2:thm:finite_l2_gain}
Given a constant $\epsilon > 0$, $H_c(\bm{\xi}, \cdot) \geq 0$, set $\alpha = \epsilon \bar{\lambda}(\bm{G}_c \bm{G}_c^\top)$ in \eqref{Chap2:eq.Controller}. Then, the controller $\Sigma_{pH}$ is $\epsilon$-output strictly passive, and has a finite $\mathcal{L}_2$-gain $\leq 1/\epsilon$. 
\end{theorem}

% \zak{ say that how the distriuted systems are passive power conserving interconnections and how it is done in the paper of Leo for sure.}

In simple words, Theorem \ref{Chap2:thm:finite_l2_gain} implies that for any choice of parameters $\bm{\theta}$, one can always choose a sufficiently large damping $\alpha$ such that the controller $\Sigma_{pH}$ is $\epsilon$-output strictly passive, and consequently, the map from $\bm{y}(t) \mapsto \bm{u}(t)$ has a finite $\mathcal{L}_2$ gain. 
Therefore, one can leverage Theorems \ref{Chap2:thm:small_gain} and \ref{Chap2:thm:finite_l2_gain} to ensure closed-loop stability in cases where the system $\Sigma_s$ is passive or has a finite $\mathcal{L}_2$ gain.
 While $\alpha$ depends on the free parameters $\bm{G}_c$, we highlight that the parametrization in Theorem \ref{Chap2:thm:finite_l2_gain} is still unconstrained since it verifies the property \eqref{Chap2:unconstrained_ppt} by design. Remark \ref{Chap2:remark_compute_alpha} in Appendix \ref{Chap2:sec:training} outlines a brief procedure for the computation of $\alpha$, when using BPTT for optimizing the cost. 

% In essence, Theorem \ref{thm:finite_l2_gain} empowers us to control a distributed system and guarantee a finite $\mathcal{L}_2$ gain by design, eliminating the need for regularization terms in $c(\bm{x}, \bm{y}, \bm{\theta})$ or the projection of weight matrices. 

% While we assume that the large-scale system $\Sigma_s$ is passive from $\bm{u}(t)$ to $\bm{y}(t)$ in our framework, 
% a sufficient condition for passivity is that all subsystems are individually passive (from $u_i(t)$ to $y_i(t)$) and the interconnection is power-conserving (e.g. skew-symmetric \cite{arcak2016networks}).
% For deeper insights into preserving passivity and 
% $\mathcal{L}_2$ gain in large-scale interconnected systems, we defer the reader to \cite{arcak2016networks}.
% Additionally, a free parametrization approach for interconnecting systems with finite $\mathcal{L}_2$ gains has been proposed in \cite{massai2023unconstrained}. 
% a parametrization of the neural-network controller \eqref{eq.Controller} such that it has finite $\ell_2$ gain. Therefore, it can be used for the applications of the small-gain theorem \cite[Theorem 2.1.1]{vanderSchaft2017} for the interconnection of multiple systems. 

\begin{remark}[Comparison with RENs] 
RENs \cite{revay2023recurrent} provide another class of models that can provide a finite $\mathcal{L}_2$ gain and are parametrized in an unconstrained way. However, differently from RENs, our parametrization allows for incorporating diverse sparsity patterns within the weight matrices. 
Moreover, our method overcomes the limitation of RENs to use only quadratic storage functions. A precise comparison with the modeling capabilities of  RENs is difficult. Nevertheless, Theorem \ref{Chap2:thm:finite_l2_gain} offers an alternative way to parametrize $\mathcal{L}_2$ operators.  
% by demonstrating robustness across a wider range of nonlinear storage functions. Hence, expanding its applicability to diverse control scenarios.
\end{remark}

\begin{remark}[Learning models with finite $\mathcal{L}_2$ gains]
Basically, the model \eqref{Chap2:eq.Controller} can be also interpreted as a Neural Ordinary Differential Equation (NODE) -- a family of continuous-depth NNs represented by dynamical systems -- \cite{NeuralODEs} with a given distributed structure and built-in $\mathcal{L}_2$ stability properties. 
Thus, besides control, it can be used for the identification of nonlinear interconnected  systems that are known to be $\mathcal{L}_2$ stable \emph{a priori}. 
While the only limitation is that the number of inputs and outputs must coincide, the possibility of embedding the $\mathcal{L}_2$ stability and the interconnection structure of the distributed system in the identification process is expected to improve the identification results. 
% Moreover, the works \cite{di2023simba,di2023stable} have demonstrated the efficacy of leveraging BPTT for system identification, albeit for linear systems only.
\end{remark}
% \begin{remark}[Communication among sub-controllers]~Note that our framework can seamlessly incorporate communication graphs among the sub-controllers without loss of generality. 
% In fact, the work \cite{furieri2022distributed} provides a systematic approach to interconnect sub-controllers while preserving the dissipativity of the closed-loop system. 
% For specific details and an example, we defer the readers to \cite[Theorem 3]{furieri2022distributed}.  
% \end{remark}

% \begin{remark}[Passivity by design]
% While achieving passivity by design for the closed-loop system is considered in \cite{furieri2022distributed}, it may not always ensure stability, especially when controlled system interacts with a passive, but else completely unknown environment.
% In fact, the converse of the passivity theorem tells us that the controlled system must be output strictly passive as seen from the interaction port of the controlled system with the environment~\cite{khong2018converse}.    
% \end{remark}

\section{Incrementally $\mathcal{L}_2$-stable Hamiltonian controllers} \label{sec:incremental_l2}

In the previous section, we demonstrated how to guarantee a finite $\mathcal{L}_2$ gain of the pH controller \eqref{Chap2:eq.Controller}. In this section, we provide formal guarantees for a finite $i\mathcal{L}_2$ gain. Following the Definition \ref{def:differential_dissipativity}, one can write the  variational dynamics for the controller $\Sigma_{pH}$ as
\begin{align}
\label{Chap2:diff_ph}
\begin{aligned}
     \delta \dot{\bm{\xi}}(t) &= (\bm{J}_c - (\alpha \bm{I} + \bm{\Lambda}))\frac{\partial^2 H_{c}(\bm{\xi},\bm{\vartheta}) }{\partial \bm{\xi}^2} \delta \bm{\xi} + \bm{G}_c \delta \bm{y}(t) \\
     \delta \bm{u}(t) &= \bm{G}_c^\top \frac{\partial^2 H_{c}(\bm{\xi},\bm{\vartheta}) }{\partial \bm{\xi}^2} \delta \bm{\xi} \;.
\end{aligned}
\end{align}
Moreover, for \eqref{Chap2:diff_ph},  differential dissipativity models the ``energy” dissipation of variations of the system trajectory. If the energy of these variations in the system trajectories decreases over time, the trajectory variation will eventually only be determined by the system input.
In order to define an incremental $\mathcal{L}_2$ gain, one can choose the supply rate as $s_\delta(\delta \bm{u}(t), \delta \bm{y}(t)) = \frac{1}{2} \gamma_\delta^2 \Vert \delta \bm{y}(t) \Vert - \frac{1}{2} \Vert \delta \bm{u}(t) \Vert$, $\gamma_\delta > 0$ \cite{vanderSchaft2017}.

% \subsection{An unconstrained parametrization of incremental $\mathcal{L}_2$ operators}

The following result, whose proof is given in Appendix \ref{Chap2:thm2:inc_proof} provides an unconstrained parametrization of pH controllers $\Sigma_{pH}$ that enjoy a finite $i\mathcal{L}_2$ gain. 
% \begin{assumption} \label{Chap2:assump.1}
% The Hamiltonian function $H_c(\bm{\xi}, \bm{\vartheta})$ of \eqref{Chap2:eq.Controller} is at least twice differentiable with respect to $\bm{\xi}(t)$  and satisfies
%     \begin{align}
%     \label{Chap2:hessian_cond}
%        0 &< c_1 \bm{I} \leq \frac{\partial^2 H_c(\bm{\xi}, \bm{\vartheta}) }{\partial \bm{\xi}^2} \leq  c_2 \bm{I}, 
%  \end{align}
% where $c_1$ and $c_2$ are positive constants.
% \end{assumption}
% ii) sufficient conditions for the differential dissipativity of the differential system \eqref{diff_ph} in terms of LMIs. These LMIs can be satisfied by design by following the steps from \cite{revay2021recurrent}.

\begin{theorem}\label{thm:incremental_l2_gain}
Given a constant $\epsilon > 0$, define $\alpha = \epsilon_\delta \bar{\lambda}(\bm{G}_c \bm{G}_c^\top)$. Moreover, let $H_c(\bm{\xi}, \bm{\vartheta})$ be at least twice differentiable with respect to its first argument and its Hessian satisfy 
    \begin{align}
    \label{Chap2:hessian_cond}
       0 &< c_1 \bm{I} \leq \frac{\partial^2 H_c(\bm{\xi}, \bm{\vartheta}) }{\partial \bm{\xi}^2} \leq  c_2 \bm{I}, 
 \end{align}
where $c_1$ and $c_2$ are positive constants.
 Then, the controller $\Sigma_{pH}$
 is $\epsilon_\delta$-output strictly incremental passive and has a finite $i\mathcal{L}_2$-gain $\leq 1/\epsilon_\delta$. 
\end{theorem}

Notably, the condition \eqref{Chap2:hessian_cond} on the Hessian of $H_c(\cdot,\cdot)$ restricts the choice to strictly convex Hamiltonian functions. One choice is to use quadratic functions to parametrize  $H_c(\bm{\xi}, \bm{\vartheta}) = \bm{\xi}^\top (\bm{X}^\top \bm{X} + \epsilon \bm{I}) \bm{\xi}$, where $\bm{X} \in \mathbb{R}^{n_c \times n_c}$ is a matrix of optimization parameters and $\epsilon > 0$ is a small positive constant. 
% However, one can also employ NN-based models for parametrizing the Hamiltonian of \eqref{Chap2:eq.Controller} for increasing the expresivity of the controller.  
However, for some complex applications, the expressivity of quadratic Hamiltonian functions may be insufficient. In this paper,  next, we provide several NN-based parametrizations that satisfy \eqref{Chap2:hessian_cond} by design.
% \begin{proposition}[Monotone neural networks by design] \label{prop.monotone_networks}
% Consider the following neural network architecture with $k$ layers
% \begin{align}
%   \bm{z}_1 &= \bm{W}_0^\top \sigma(\bm{W}_0 \bm{\xi} + \bm{b}_0) \label{eq.monotone_arch} \\ 
%   \bm{z}_{i+1} &= \bm{W}_{i}^\top \sigma(\bm{W}_i \bm{z}_i + \bm{b}_i), \ i= 1, \cdots, k-1 \label{eq.monotone_arch2}\\ 
%   \bm{\pi}(\bm{\xi}) &= \bm{z}_k \;,\nonumber
% \end{align} 
% where  $\sigma(\cdot)$ is a monotonically increasing activation function and satisfies $0 \leq \sigma'(\cdot) \leq S$, $\bm{z}_i$  are the hidden states, and $\theta := \{\bm{W}_i, \bm{b}_i \}$ are real-valued trainable parameters. Then, the map $\pi(\bm{\xi})$ 
% is monotonic with respect to $\bm{\xi}$ for all trainable parameters $\theta$.
% \end{proposition}

% Motivation for monotone neural networks! \cite{runje2022constrained, liu2020certified}.

\begin{remark}[Monotone networks]
   The condition \eqref{Chap2:hessian_cond} can be satisfied by parametrizing the gradient of the Hamiltonians $H_i({\xi}_i, {\vartheta}_i)$ as
    \begin{equation} \label{eq.param_mon}
       \frac{\partial H({\xi}_i, {\vartheta}_i)}{\partial {\xi}_i}  = {z}_1({\xi}_i)  + \epsilon {\xi}_i \; ,
    \end{equation}
where ${z}_1({\xi}) = {W}_0^\top \sigma_m({W}_0 {\xi}_i + {b}_0)$ is  a single-layered NN that is monotone\footnote{In functional analysis, on a topological vector space 
$\Xi$, a  non-linear operator $z_1: \Xi \rightarrow \Xi$ is said to be a monotone operator if $\langle z_1({\xi}_1) - z_1({\xi}_2), {\xi}_1 - {\xi}_2  \rangle \geq 0$.} with respect to ${\xi}$ for all ${W}_0 \in \mathbb{R}^{n_c \times n_c}$, $\sigma_m(\cdot)$ is a monotonically increasing activation function whose derivative $\sigma_m'(\cdot)$ satisfies $0 \leq \sigma_m'(\cdot) \leq \bar{S}$, and $\epsilon > 0$ is a small positive constant. Recall that a function is convex if and only if the gradient of the function is monotone \cite[Proposition B.9]{foucart2013mathematical} and in view of \eqref{eq.param_mon}, the Hamiltonian function $H_i({\xi}_i, {\vartheta}_i) = \int ({z}_1({\xi}_i)  + \epsilon {\xi}_i) d {\xi}_i$ is convex. Moreover, the Hessian of $H_i({\xi}_i, {\vartheta}_i)$ can be computed as 
% \begin{equation*}
%     \frac{\partial^2 H(\bm{\xi}, \theta) }{\partial \bm{\xi}^2} = \bm{W}_{k-1}^\top \prod_{i = 0}^{k-1} \bm{D}_i(\cdot) \bm{W}_0 + \epsilon \bm{I}\;.
% \end{equation*}
\begin{equation*}
    \frac{\partial^2 H_i({\xi}_i, {\vartheta}_i) }{\partial {\xi}^2_i} = {W}_{0}^\top  {D}(\cdot) {W}_0 + \epsilon {I}_{q_i}\;.
\end{equation*}
Since the diagonal matrices ${D}(\cdot) := \operatorname{diag}(\sigma'({W}_0 (\cdot) + {b}_0)$ and verify $0 \leq {D}(\cdot) \leq \bar{S}$, the constants $c_1$ and $c_2$ in \eqref{Chap2:hessian_cond} can be easily computed as 
\begin{equation*}
    c_1 = \epsilon, \ c_2 = \bar{S} \bar{\lambda}({W}_{0}^\top {W}_0) + \epsilon\;.
\end{equation*}
\end{remark}

Note that the assumption that activation functions $\sigma(\cdot)$ are monotonic is satisfied by almost every nonlinear activation function used in practice, e.g., $\operatorname{tanh}(\cdot)$, $\operatorname{SmoothReLU}(\cdot)$, $\operatorname{SmoothLeakyReLU}(\cdot)$, and $\operatorname{sigmoid}(\cdot)$. 

\begin{remark}[ICNNs]
    In Theorem \ref{thm:incremental_l2_gain} one can parametrize the  Hamiltonian functions $H_i({\xi}_i,{\vartheta}_i)$ by ICNNs \cite{amos2017input} given in Remark \ref{Chap2:remark_hamiltonian}.   
     Since  \eqref{Chap2:eq:icnns} is strictly convex in ${\xi}_i$, the Hessian can be bounded by the constants $c_1 {I}_{q_i}$ and $c_2 {I}_{q_i}$. These constants can be calculated using the \emph{autograd} library of Pytorch \cite{pytorch}, for example. 
% \zak{Closed forms of $c_1$ and $c_2$ are difficult to obtain!}
\end{remark}

Furthermore, the strict convexity of $H_c(\bm{\xi}, \bm{\vartheta}) = \sum_{i \in \mathcal{V}_s} H_i(\xi_i,\vartheta_i)$ follows from the fact that the sum of strictly convex functions is also strictly convex.

\section{Dissipation-preserving discretization schemes}
\label{Chap2:discretization_controllers}
Given a continuous-time passive system,  its discretized counterpart may not enjoy the same properties, as highlighted in \cite{xia2014passivity}, and \cite{joo2019preserving}. For instance, traditional discretization methods, such as zero-order hold, do not preserve passivity.\footnote{We defer the reader to \cite{martinelli2023interconnection} for the formal definitions of the discrete-time notion of passivity.} Several researchers have focused on geometric discretization schemes based on variational integration theory, including (Semi-) Implicit Euler, leap-frog, and Verlet integration methods \cite{hairer2006geometric}. While they can conserve structural properties such as \emph{symplecticity}, and long-term energy stability, they do not preserve the passivity, finite $\mathcal{L}_2$, and  $i\mathcal{L}_2$ gains. Furthermore,  discretized systems stemming from these methods may become unstable for a large sampling period as shown in \cite{joo2019preserving} and also illustrated in Appendix  \ref{subsec:disspation_preserving}. In this section, we leverage state-of-the-art discrete gradient methods \cite{ehrhardt2018geometric} to preserve the dissipative properties of pH controllers \eqref{Chap2:eq.Controller} after discretization.
To this end, we define discretizations of the gradient $\nabla H_c(\bm{\xi}) \equiv \frac{\partial H_c(\bm{\xi})}{\partial \bm{\xi}}$
to define a class of numerical integrators that exactly preserve the first integral $H_c(\cdot)$.

\begin{definition}[\cite{gonzalez1996time}] \label{Chap2:def:discrete_gradients}
Let $H : \mathbb{R}^n \mapsto \mathbb{R}$ be a differentiable function. Then $\bar{\nabla} H : \mathbb{R}^{2n} \mapsto \mathbb{R}^n$ is a discrete gradient of $H$ if it is continuous and satisfies
\begin{align*}
    \bar{\nabla} H(\bm{x},\tilde{\bm{x}})^\top (\tilde{\bm{x}}- \bm{x})= H(\tilde{\bm{x}})- H(\bm{x}), \forall \bm{x},\tilde{\bm{x}} \in \mathbb{R}^n, \\
     \bar{\nabla} H(\bm{x},\bm{x}) = {\nabla}H(\bm{x}), \forall \bm{x} \in \mathbb{R}^n .
\end{align*}
\end{definition}

Some well-known examples of discrete gradients are:
\begin{itemize}
    \item the mean value (or averaged) discrete gradient \cite{harten1983upstream}
    \begin{equation*}
        \bar{\nabla}H(\bm{x}, \tilde{\bm{x}}) := \int_{0}^1 \nabla H((1-s)\bm{x} + s \tilde{\bm{x}})ds \;;
    \end{equation*}
    \item the midpoint (or Gonzalez) discrete gradient \cite{gonzalez1996time}
    \begin{align*}
            &\bar{\nabla}H(\bm{x}, \tilde{\bm{x}}) :=  \nabla H (\frac{1}{2} ( \bm{x} + \tilde{\bm{x}})) \\ & + \frac{H(\tilde{\bm{x}}) - H(\bm{x}) -  \nabla H (\frac{1}{2} ( \bm{x} + \tilde{\bm{x}}))^\top (\tilde{\bm{x}} - \bm{x})}{\vert \tilde{\bm{x}} - \bm{x}\vert^2} (\tilde{\bm{x}} - \bm{x})
    \end{align*}
    for $\tilde{\bm{x}} \neq \bm{x}$;
    \item the coordinate increment discrete gradient (Itoh-Abe)~\cite{itoh1988hamiltonian} with each component given by 
    \begin{align*}
        \bar{\nabla}H(\bm{x}, \tilde{\bm{x}})_i := \frac{H(\tilde{\bm{x}}_1, \dots, \tilde{\bm{x}}_i, \bm{x}_{i+1}, \dots, \bm{x}_n)}{\tilde{\bm{x}}_i - \bm{x}_i} \\
        - \frac{H(\tilde{\bm{x}}_1, \dots, \tilde{\bm{x}}_{i-1}, \bm{x}_{i}, \dots, \bm{x}_n)}{\tilde{\bm{x}}_i - \bm{x}_i}
    \end{align*}
    for $1 \leq i \leq n$ and $\tilde{\bm{x}}_i \neq \bm{x}_i$.
    
\end{itemize}
For more details on these methods, see \cite{ehrhardt2018geometric}.
% In the following, we leverage discrete gradient methods (Definition \ref{def:discrete_gradients}) to preserve the dissipative properties of the continuous-time controllers \eqref{eq.Controller} stemming fro  via   finite $\ell_2$ gain.  

Consider the following numerical discretization of controller \eqref{Chap2:eq.Controller}

\begin{equation}
\Sigma_{dpH}: \begin{cases}
\begin{aligned}
      \frac{\bm{\xi}_{k+1} - \bm{\xi}_k}{h_{\Delta}} &= (\bm{J}_c - (\alpha \bm{I} + \bm{\Lambda})) \bar{\nabla}H_c(\bm{\xi}_{k+1}, \bm{\xi}_k) \\
      & \quad + \bm{G}_c \bm{y}_k \\ 
    \bm{u}_k &= \bm{G}^\top_c \bar{\nabla}H_c(\bm{\xi}_{k+1}, \bm{\xi}_k) \;,  
\end{aligned}\label{Chap2:eq:discrete_cont}
\end{cases}
\end{equation}
% \begin{align} \label{Chap2:eq:discrete_cont}
%    \label{Chap2:eq:discrete_cont2}
% \end{align}
where $h_{\Delta}>0$ is the sampling time, and $\bm{\xi}_k, \bm{y}_k, \bm{u}_k$ corresponds to $\bm{\xi}(kh_{\Delta}), \bm{y}(kh_{\Delta})$, and $\bm{u}(kh_{\Delta})$, respectively. Let us choose $\alpha = \epsilon \bar{\lambda}(\bm{G}_c \bm{G}_c^\top)$ as in Theorem \ref{Chap2:thm:finite_l2_gain} so that the continuous-time counterpart \eqref{Chap2:eq.Controller} has a finite $\mathcal{L}_2$ gain. Our goal is to prove that for this choice of dissipation constant $\alpha$, the controller \eqref{Chap2:eq:discrete_cont} preserves the $\mathcal{L}_2$ gain. To this aim, the following result, whose proof is provided in Appendix \ref{Chap2:proof_of_thm_5} shows that the discrete-time controller is also $\epsilon$-output  strictly passive. 

\begin{theorem} \label{Chap2:thm:discrete_L2_gain}

Given a constant $\epsilon > 0$, let $\alpha = \epsilon \bar{\lambda}(\bm{G}_c \bm{G}_c^\top)$, and $H_c(\bm{\xi}, \cdot) \geq 0$. Then, the discrete-time controller $\Sigma_{dpH}$
 is $\epsilon$-output strictly passive, and
   has a finite $\mathcal{L}_2$-gain $\leq 1/\epsilon$. 

\end{theorem}

Unfortunately, the discrete-time dynamics \eqref{Chap2:eq:discrete_cont} is implicit, that is, $\bm{\xi}_{k+1}$ appear on both sides of the dynamics, and the output equation $ \bm{u}_k = \bm{G}^\top_c \bar{\nabla}H_c(\bm{\xi}_{k+1}, \bm{\xi}_k) $ is non-causal. Note that having implicit forms in the dissipation-preserving discretization of pH systems is quite common \cite{macchelli2023control}. Moreover, if $H_c(\cdot,\cdot)$ is parametrized via a NN, the discrete-time dynamics \eqref{Chap2:eq:discrete_cont} defines the forward equation of a well-known class of NNs called \emph{implicit NNs}-see \cite{jafarpour2021robust, smith2022physics, jafarpour2022robustness, winston2020monotone, revay2023recurrent}, for example. 
The main challenge of employing implicit NNs in practice is to guarantee the existence and uniqueness of the solution of the forward equation. In our case, this amounts to show that the following implicit equation has a unique solution $\bm{\xi}_{k+1} \in \Xi$ for all $\bm{\xi}_k \in \Xi$ and $\bm{y}_k \in \mathcal{Y}_c$
\begin{align} 
 \bm{\xi}_{k+1}  = \bm{\xi}_k &+ h_{\Delta}(\bm{J}_c - (\alpha \bm{I} + \bm{\Lambda})) \bar{\nabla}H_c(\bm{\xi}_{k+1}, \bm{\xi}_k) \nonumber \\ &+ h_{\Delta}\bm{G}_c \bm{y}_k \;. \label{Chap2:eq:implicit} 
\end{align}
Sometimes, such implicit expressions can be easily solved for $\bm{\xi}_{k+1}$, e.g. in the linear case \cite{moreschini2020stabilization}.
% Moreover, several works propose to enforce constraints on the optimization parameters of implicit NNs such that the forward equation is contractive, thereby guaranteeing the uniqueness of the solution \cite{jafarpour2021robust, winston2020monotone}. 
% However, since  \eqref{Chap2:eq:implicit} is nonlinear and the term $\bar{\nabla}H_c(\bm{\xi}_{k+1}, \bm{\xi}_k)$ is non-separable,\footnote{Non-separable refers to the quantities for which the terms that only depend on $\bm{\xi}_{k+1}$, or $\bm{\xi}_k$ cannot be separated.} enforcing contraction by constraining optimization parameters is not straightforward and also undermines the principle of unconstrained parametrization. 
Nevertheless, a sufficient condition on the existence and uniqueness of the solution of \eqref{Chap2:eq:implicit} is provided in \cite{macchelli2023control}. This condition is based on the uniform boundness of $\nabla H_c(\bm{\xi})$, and the choice of discrete gradient used. The technical details have been omitted here.

\begin{remark}[Finite $i\mathcal{L}_2$ gain]
    By selecting the dissipation $\alpha \geq \epsilon_\delta \bar{\lambda}(\bm{G}_c \bm{G}_c^\top)$ in the pH controller \eqref{Chap2:eq.Controller} and the Hamiltonian function $H_c$ such that it satisfies condition \eqref{Chap2:hessian_cond}, the implicit discrete-time controller \eqref{Chap2:eq:discrete_cont} has a finite $i\mathcal{L}_2$ gain. 
    This can be proved similarly as for Theorem \ref{Chap2:thm:discrete_L2_gain}. 
\end{remark}

% \textcolor{red}{Preserving the differential input-output stability described in Theorem \ref{Chap2:thm.diff_ext_stable} using a time-varying Hamiltonian function through discrete gradient methods is not straightforward. This difficulty arises because Definition \ref{Chap2:def:discrete_gradients} for discrete gradients is only applicable to time-invariant Hamiltonian functions. While some approaches treat time as an additional state and propose extended Hamiltonians (see, for example, \cite{de2011symplectic} or \cite{hairer2006geometric}), the study of Hamiltonian systems with dissipation in the extended phase space is beyond the scope of this paper.}

\section{Experiments} \label{sec:experiments}

In this section, we showcase the application of the proposed pH-based control framework via experiments. Specifically, we consider consensus control of non-holonomic wheeled robots, and power sharing and weighted voltage regulation in a DC microgrid. 
% Additionally, we present a simple example highlighting the importance of discrete gradient methods for preserving the dissipative properties of NN controllers in real-world implementation. 
The readers are also referred to \cite{zakwan2024neural} for an additional example about the synchronization of Kuramoto oscillators using the pH controllers \eqref{Chap2:eq.Controller} endowed with a finite $\mathcal{L}_2$ gain.

\subsection{Consensus of non-holonomic wheeled robots}
\label{Chap2:example1}
\begin{figure}
    \centering
    \includegraphics[width = 0.5\linewidth]{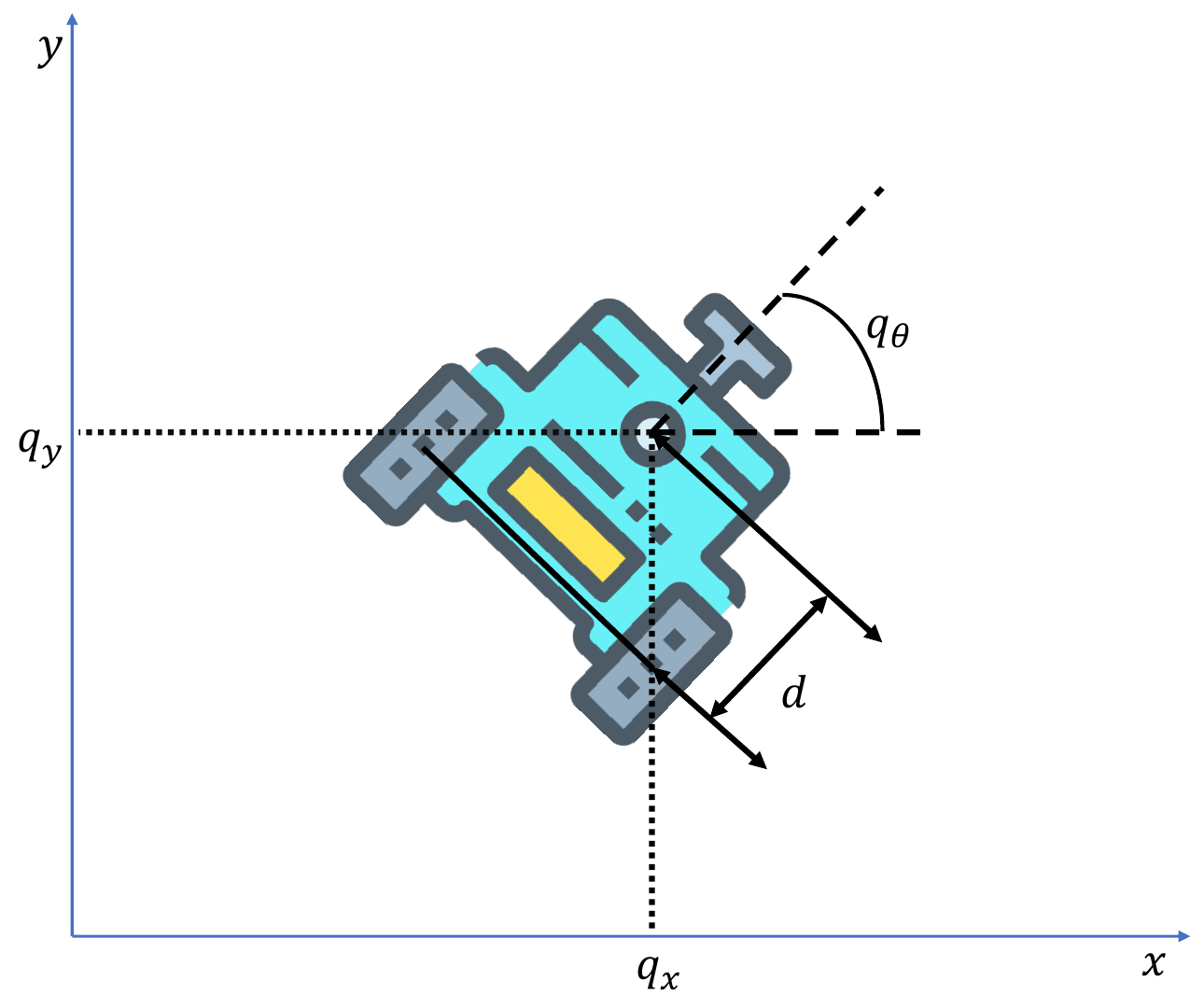}
    \caption{The mobile wheeled robot used in the experiment. The triple $(q_x, q_y, q_\theta)$ denotes the position and the orientation of the robot, and $d$ is the distance from the center to the wheels.}
    \label{fig:my_robo}
\end{figure}

We illustrate the efficacy of the pH controller \eqref{Chap2:eq.Controller} by considering a consensus problem with collision avoidance for a swarm of $N$ non-holonomic physically decoupled agents such as wheeled mobile robots modeled as  \cite{tsolakis2021distributed}, \cite{gimenez2015port}:
\begin{equation}
\label{swarm_exp}
        \begin{array}{lll}
             \dot{ {x}}_i(t)
              = 
             (J_s({x}_i) - R_s) 
        
             {\partial {H}({{{x}_i}}) \over \partial{ {{x}_i}} } +  {{G}(x_i)} { {{u}_i}}(t)\; \\
             {{y}_i}(t) =  {{G}(x_i)}^\top  {\partial {H}({ {{x}_i}}) \over \partial{ {{x}_i}} } \;,
        \end{array}
\end{equation}
where ${{x}}_i = [{q}_i^\top, {{p}}_i^\top]^\top \in \mathbb{R}^{n}$, and ${q}_i = [q_{x,i}, q_{y,i}, q_{\theta,i}]$ are generalized coordinates and $p_i = [p_{v,i}, p_{\omega,i}]$ are transformed generalized momenta for $i^{th}$ robot as depicted in Fig. \ref{fig:my_robo}.
Moreover, the matrix-valued functions in \eqref{swarm_exp} are given as:
% ${J}_s = \left[
%           \begin{matrix}
%                     {0}_{3 \times 3}  & {S}({q}) \\ 
%                     -{S}^\top({q})  & {C}({q},{p})
%           \end{matrix}
%           \right]$, $ \bar{\bm{G}} = \left[ 
%           \begin{matrix}
%           \bm{0}_{3 \times 2}  \\ {\tilde{\bm{G}}(\bm{q})}
%           \end{matrix}
%           \right]$, with  
 \begin{equation}
    \begin{array}{rll}
    {J}_s(x_i) &=& \left[
          \begin{matrix}
                    {0}_{3 \times 3}  & {S}({x_i}) \\ 
                    -{S}^\top({x_i})  & {C}({x_i})
          \end{matrix}
          \right],  \\
         {S}({x_i}) &=& \left[ 
         \begin{matrix}
         \cos{q_{\theta,i}} & - d \sin{q_{\theta,i}} \\ 
         \sin{q_{\theta,i}} & d \cos{q_{\theta,i}}  \\ 
         0 & 1
         \end{matrix}
         \right],  \\
     {C}(x_i) &=& \left[ 
        \begin{matrix}
        0 & {md \over md^2 + \Omega_m} p_{\omega,i}  \\ 
        - {md \over md^2 + \Omega_m} p_{\omega,i} & 0
        \end{matrix}
        \right], \\
        R_s &=& 0.05 I_5, \  G(x_i) = \left[ 
          \begin{matrix}
        {0}_{3 \times 2}  \\ I_{2}
          \end{matrix}
          \right] \;.
    \end{array}
\end{equation}        
In \eqref{swarm_exp}, $H_i(x_i) = \frac{1}{2} p^\top M p, M = \texttt{blkdiag}(m, \Omega_m + md^2)$ is the kinetic energy of an individual robot, where $m= 6.0 Kg$, $\Omega_m = 0.1 Kg m^2$, and $d = 0.2 m$. For further details about model \eqref{swarm_exp}, we defer the reader to \cite{tsolakis2021distributed}, and \cite{gimenez2015port}. 

Our goal is to minimize the following loss function:
\begin{align}
\label{optimal_loss}
   \int_0^T (\ell_1(\bm{x}(t)) + \ell_2(\bm{x}(t)) )dt\;
\end{align}
with the consensus-promoting term
\begin{equation*}
    \ell_1(\bm{x}(t)) = {1 \over N^2} \sum_{i = 1}^N \sum_{j = 1}^N ||v_i - v_j|| \; ,
\end{equation*}
and the collision-avoidance term 

\begin{equation*}
 \ell_2(\bm{x}(t)) = \begin{cases}
\begin{aligned}
      {1 \over N^2}\sum_{i = 1}^{N-1} \sum_{j = i+1}^N {1 \over (r_{ij} + \epsilon)^2}, \ \text{if} \ r_{ij} < \bar{r} \\  0 \ \text{otherwise}\; , 
\end{aligned}
\end{cases}
\end{equation*}

where $v_i$ is the forward velocity of agent $i$, $r_{ij}$ is the relative distance between an agent $j$ and an agent $i$, $\bar{r} > 0$ is a constant that defines the collision-avoidance distance, and $\epsilon>0$ is a small positive constant for numerical stability.  Since the radius of each robot is 
$0.2m$, a relative distance of 
$r_{ij} < 0.5m$ (from the center of the $i^{th}$ robot to the center of the 
$j^{th}$ robot) will denote a collision, therefore, in this experiment, we set $\bar{r} = 0.8m$. To compute the loss function \eqref{optimal_loss}, the relative distance $r_{ij}$ can be calculated as $r_{ij} = \sqrt{(q_{x,i} - q_{x,j})^2 + (q_{y,i} - q_{y,j})^2}$ and $v_i$ in $\ell_1(\bm{x}(t))$ is computed as $p_{v,i}/m, \ \forall i \in \mathcal{V}_s$. 
% \begin{equation}
% \label{optimal_loss}
%  \int_0^T (\bm{x}-\bm{x}_f)^\top \bm{Q}(\bm{x}-\bm{x}_f) + \bm{u}^\top \bm{R} \bm{u} \ dt\;, ~ \bm{x}(0) = \bm{x}_0
% \end{equation}
% for some given positive semi definite matrices $\bm{Q} \in \mathbb{R}^{n \times n}$, $\bm{R} \in \mathbb{R}^{m \times m}$ and the time-horizon $T > 0$.
% WHat are non-holomic agents and why they are important

% Non-holonomic mechanical systems encompass a large class of practically interesting robotic structures, such as wheeled mobile robots, space manipulators, and multi-fingered robot hands.
% The dynamics of a non-holonomic system are defined by sets of equations of motion and non-integrable differential equations known as non-holonomic constraints. These constraints lead to the coupling among the generalized velocities of the system and make it cumbersome to control \cite{tsolakis2021distributed}. 
It is well known that dynamical systems subject to non-holonomic constraints violate \emph{Brockett's necessary condition} for asymptotic stabilizability, meaning they cannot be stabilized using a continuous state-feedback policy~\cite[Section 6.4]{vanderSchaft2017}. This motivates the use of time-varying and discontinuous controllers, such as those parametrized by pH controllers \eqref{Chap2:eq.Controller} with time-varying parameters for $t \leq T$, while minimizing the nonlinear objective function \eqref{optimal_loss}. For $t > T$, we freeze the parameters of the controller, as discussed in Remark \ref{Chap2:remark:time_varying_weights}.

\begin{figure}
    \centering
    \includegraphics[width=\linewidth]{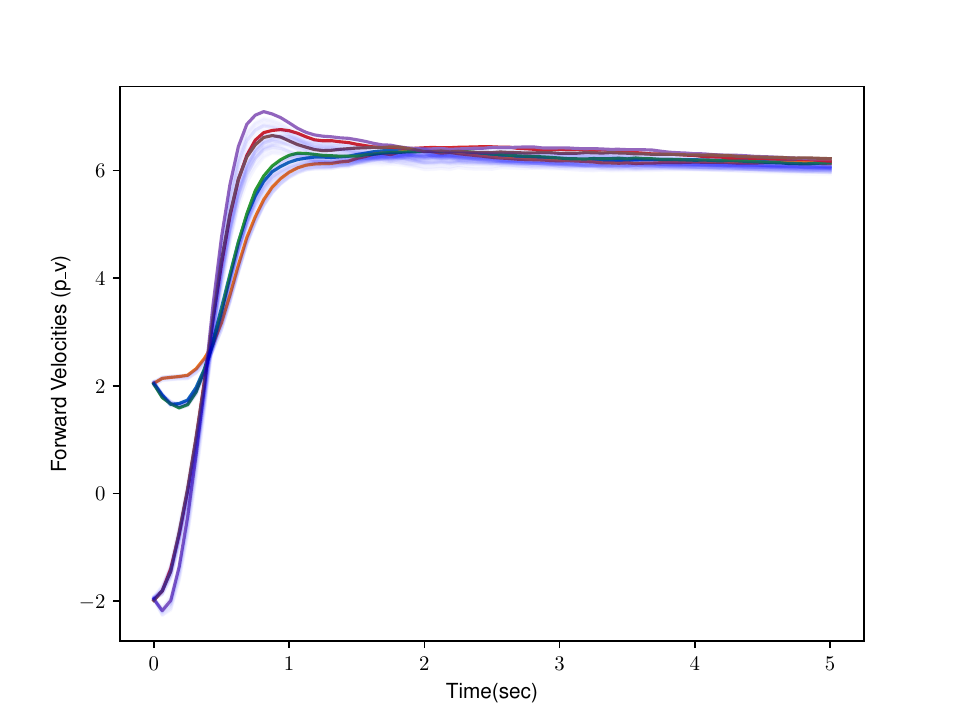}
    \caption{Forward velocities of the swarm demonstrating 
consensus. The translucent blue color lines represent the evolution of trajectories starting from the initial conditions sampled from a ball with radius $0.05$ centered at the nominal condition for each agent.}
    \label{Chap2:fig:consensus}
\end{figure}
% \begin{figure}
%     \centering
%     \includegraphics[width=\linewidth]{Figs/control_distributed.pdf}
%     \caption{NN controller energies.}
%     \label{Chap2:fig:control_energies}
% \end{figure}
\begin{figure}
    \centering
    \includegraphics[width=\linewidth]{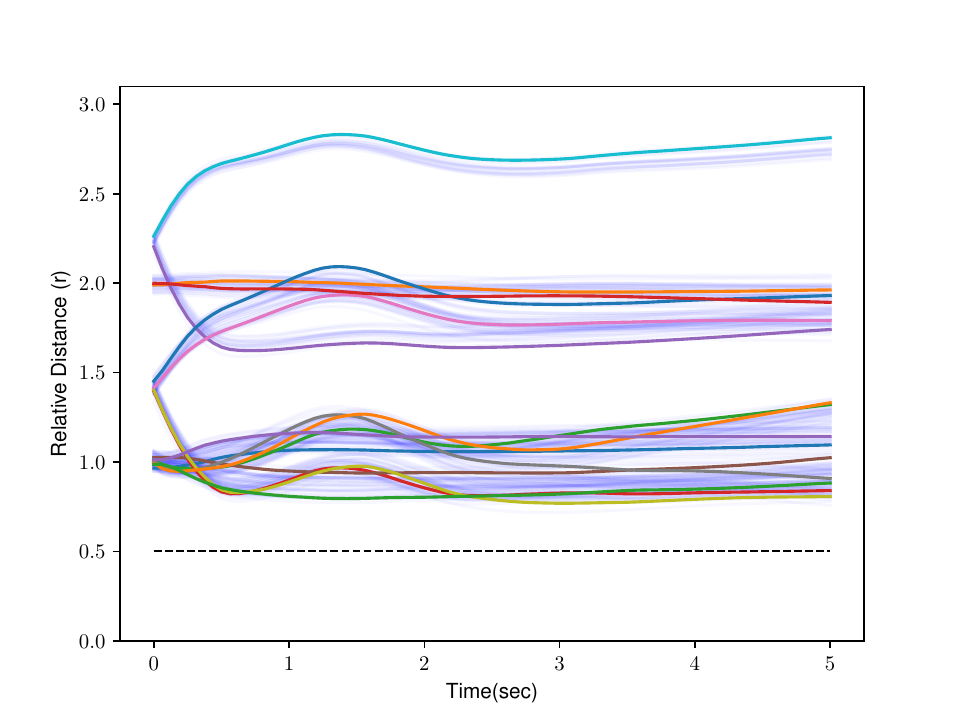}
    \caption{Relative distance $r_{ij}$ among the agents with collision
avoidance. Note that $r_{ij} < 0.5$ indicates a collision. The translucent blue color lines represent the evolution of trajectories starting from the initial conditions sampled from a ball with radius $0.05$ centered at the nominal condition for each agent.}
    \label{Chap2:fig:relative_distances}
\end{figure}

 To solve the OCP with loss function \eqref{optimal_loss}, we trained the distributed pH controller \eqref{Chap2:eq.Controller}, 
where we set $\bm{J}_c = \texttt{blkdiag}(J_i)$, with  
${J}_i = I_2 \otimes \begin{bmatrix}
  0 & 1 \\ -1 & 0
\end{bmatrix}$,\footnote{Here, $\otimes$ denotes the standard Kronecker product.} $\alpha = 0.125 \bar{\lambda}(\bm{G}_c \bm{G}_c^\top)$ as in Theorem \ref{Chap2:thm:finite_l2_gain}, $\bm{\Lambda} \equiv 0$, and $\bm{G}_c = \texttt{blkSparse}(\mathcal{P})$, where ${\mathcal{P}} \in \mathbb{R}^{N \times N}$ is the adjacency matrix. To have more flexibility during the finite horizon $T$, we choose a time-varying Hamiltonian function for $i$-th controller, i.e.  $H_c({\xi}_i,t) = \log[\cosh({ K}_i(t){\xi}_i + {b}_i(t))]^\top \mathds{1}$ for $t \in [0, T]$, where ${K}_i: \mathbb{R} \rightarrow \mathbb{R}^{4 \times 4}$ and ${b}_i: \mathbb{R} \rightarrow \mathbb{R}^4$ are piecewise constant optimization parameters.
In this investigation, we consider  $N=6$ agents communicating over a network with an undirected cyclic graph topology, i.e. whose adjacency matrix is given by  
\begin{equation*}
    \begin{array}{ll}
         \mathcal{P} = \left[ 
         \begin{matrix}
         2 & -1 & 0 & 0 & 0 & -1 \\ 
         -1 & 2 & -1 & 0 & 0 & 0\\ 
         0 & -1 & 2 & -1 & 0 & 0\\ 
         0 & 0 & -1 & 2 & -1 & 0 \\  
         0 & 0 & 0 & -1 & 2 & -1  \\
         -1 & 0 & 0 & 0  & -1 & 2 
         \end{matrix}
         \right] \; .
    \end{array}%
\end{equation*}

Then, we optimize the parameters $\bm{\theta} = \{ \bm{K}, \bm{b}, \bm{G}_c\}$ using BPTT in PyTorch \cite{pytorch}. Particularly, we choose $S=20$ initial conditions sampled uniformly in a ball of radius $0.05$ centered at nominal initial condition $\bm{x}_0$ for each agent. The whole training procedure takes almost $10$ mins on a Bizon ZX5000 G2 workstation. For more details, we defer the reader to Appendix \ref{Chap2:sec:training}.
% Our goal is to drive the wheeled robot from an initial position $\bm{q}_0 = [1,1,0]^\top$ to a final position $\bm{q}_f = [4,4,q_\theta]^\top$ ($q_\theta$ is an arbitrary orientation of the agent) while minimizing the finite-horizon loss function \eqref{optimal_loss} with $T = 5$ sec. We choose $\bm{Q} = diag(5,5,0,0.2,0)$, $ \bm{R} = diag(0.1,0.1)$ for optimization. 

After optimization, we simulate the closed-loop system. Fig. \ref{Chap2:fig:consensus}  demonstrates the consensus among all agents in terms of forward velocities, while Fig. \ref{Chap2:fig:relative_distances} displays the relative distances among the agents initialized from various initial conditions. Since the relative distances $r_{ij} > 0.5, \forall i,j \in \mathcal{V}_s$, the pH controller has effectively achieved collision avoidance. 
Moreover, in both figures, the translucent blue trajectories demonstrate that the pH controller managed to achieve the goals for new test points sampled near the nominal initial conditions, indicating a degree of robustness to out-of-sample distributions.

\subsection{Power sharing and averaged voltage regulation in DC microgrids}
\label{Chap2:sec:example2}

\begin{figure}
    \centering\includegraphics[width=\linewidth]{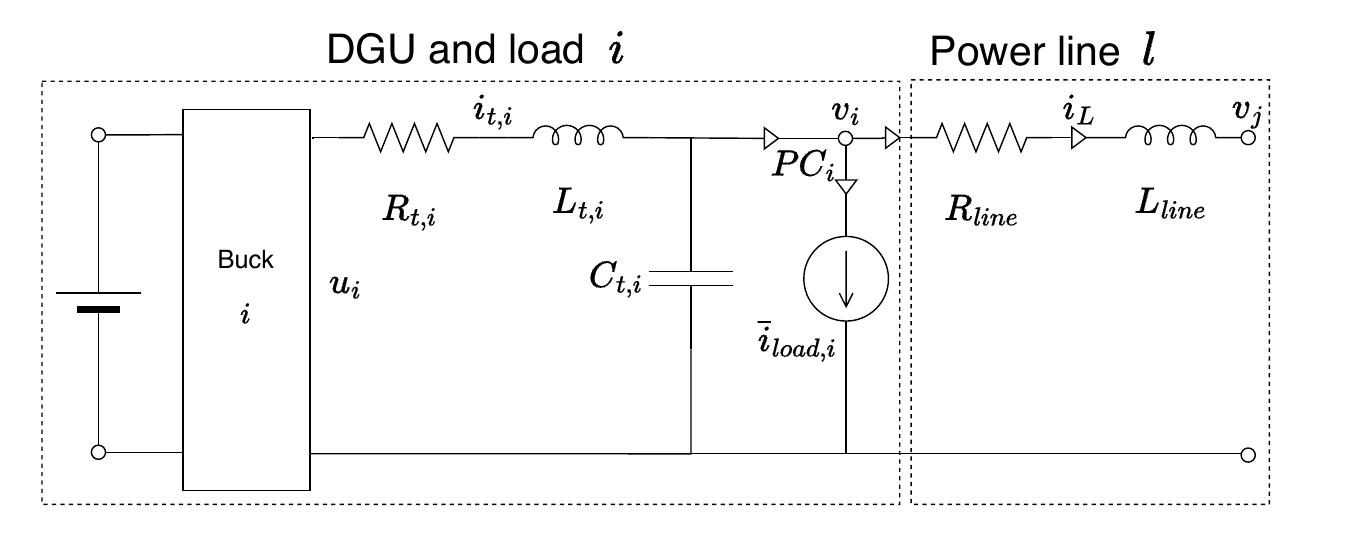}
    \caption{
The electric scheme of $i^{th}$ DGU and the collocated load.}
    \label{fig:enter-label:circuit}
\end{figure}

\begin{figure}
    \centering
    \includegraphics[width=\linewidth]{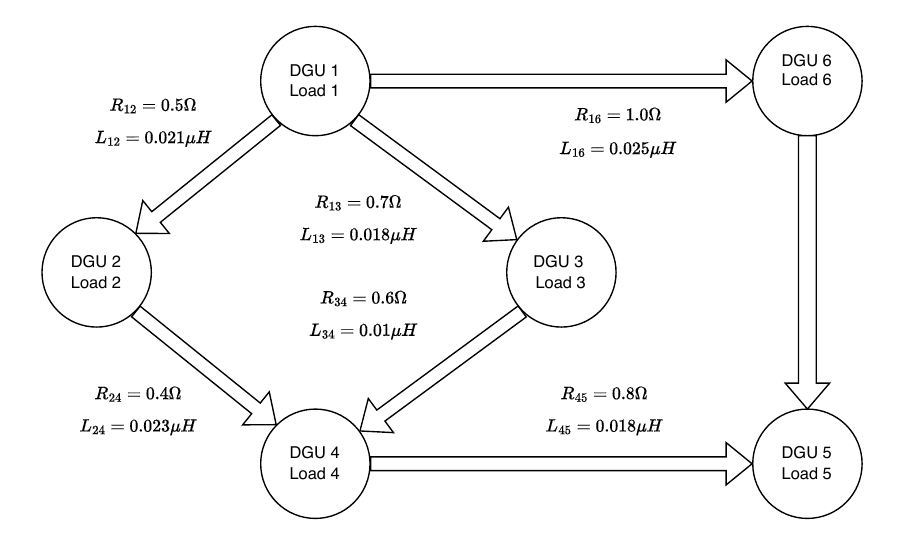}
    \caption{Representation diagram of the DC microgrid with line parameters. The arrows show the reference system for positive currents. The undirected communication topology among DGUs' controllers is given by the same graph by ignoring the arrows.}
    \label{Chap2:fig.DGUs}
\end{figure}

In this experiment, we explore weighted power sharing and averaged voltage regulation in DC microgrids. The model of a DC microgrid consists of $N$ Distributed Generation Units (DGUs) connected by $M$ resistive-inductive (RL) power lines \cite{nahata2021consensus}.  An electrical circuit diagram of a typical DGU with its load is provided in Fig. \ref{fig:enter-label:circuit}.
 The overall DC microgrid can be described as follows~\cite{nahata2021consensus}:
\begin{align*}
\Sigma_{dmG} := \begin{cases}
    \bm{C}_t \dot{\bm{v}}= \bm{i}_t + \mathcal{B} \bm{i}_L - \bm{i}_{load} \\
    \bm{L}_t \dot{\bm{i}}_t = - \bm{v} - \bm{R}_t \bm{i}_t + \bm{u}(t) \\ 
    \bm{L}_{line} \dot{\bm{i}}_L = - \mathcal{B} \bm{v} - \bm{R}_{line} \bm{i}_L \\ 
    \bm{y}(t) = \bm{i}_t\;,
\end{cases}
\end{align*}
% where $\bm{i}_t, \bm{v} \in \mathbb{R}^N$ are the stacked filter (generator) currents, voltages at point of connections, and $\bm{I}_L \in \mathbb{R}^M$ is the vector of line currents, respectively. The control input $\bm{u}(t)$  is the command to the dc-dc buck converter, and $\bm{i}_{load}$ is the current drawn by the loads, in this example, for the sake of simplicity, we consider only ZI loads: $\bm{i}_{load} = \bm{Y}_{load} \bm{v} + \bar{\bm{i}}_{load}$. 
%  Furthermore, $\bm{C}_t, \bm{L}_t, \bm{Y}_{load}, \bm{R}_t \in \mathbb{R}^{N \times N}$ and $\bm{R}_{line}, \bm{L}_{line} \in  \mathbb{R}^{M \times M}$ are positive definite diagonal matrices, e.g. $\bm{C}_t = \texttt{diag}(C_1, \dots, C_N )$, where $C_i$ are the capacitance (lumped with the line capacitances) of the DC converters, $\bm{L}_t$, $\bm{R}_t$ are the internal inductances and resistances. Finally, $\bm{R}_{line}$, and $\bm{L}_{line}$ are the inductances and resistances of power lines connecting the DGUs. We assume the electrical and communication interconnections are same, and is given by 
where $\bm{i}_t$ and $\bm{v} \in \mathbb{R}^N$ represent the stacked filter (generator) currents and the voltages at the points of connection, respectively, while $\bm{I}_L \in \mathbb{R}^M$ denotes the vector of line currents. The control input $\bm{u}(t)$ is the stacked vector of all the command signals to DC-DC Buck converters, and $\bm{i}_{load}$ collects all the currents drawn by the loads collocated with DGUs. For the sake of this experiment, we assume that the electrical and communication interconnections are identical and are given by the graph in Fig. \ref{Chap2:fig.DGUs} whose incidence matrix is:
 \begin{align}
    \mathcal{B}=  \begin{bmatrix}
        1 & 0 & 1 & 0 & 0 & 1 &  0 \\
        -1 & 1 & 0 & 0 & 0 & 0 &  0  \\
        0&  0& -1& 1& 0& 0 & 0   \\
        0 & -1&  0& -1&  1&  0 & 0  \\
        0&  0& 0&  0&  -1& 0 & -1 \\
        0& 0& 0& 0& 0& -1 & 1
    \end{bmatrix}  \; .            
 \end{align}
The graph adjacency matrix is therefore defined as $\mathcal{P}_e = \mathcal{B}  \Gamma_{w} \mathcal{B}^\top
$, where $\Gamma_{w}$ are the weights of the edges. All the parameters of the DC microgrid considered in this experiment are given in Table \ref{Chap2:params_grid}, while the interconnection of the DGUs and corresponding line parameters $R_{ij}, L_{ij}$ are provided in Fig. \ref{Chap2:fig.DGUs}. 
We consider loads modeled by an impedance and a constant current, i.e.  $\bm{i}_{load} = \bm{Y}_{load} \bm{v} + \bar{\bm{i}}_{load}$.
Additionally, $\bm{C}_t$, $\bm{L}_t$, $\bm{Y}_{load}$, and $\bm{R}_t \in \mathbb{R}^{N \times N}$, as well as $\bm{R}_{line}$ and $\bm{L}_{line} \in \mathbb{R}^{M \times M}$, are positive definite diagonal matrices. For example, $\bm{C}_t = \texttt{diag}(C_1, \dots, C_N )$, where $C_i$ represents the capacitance (combined with line capacitances) of the DC converters, and $\bm{L}_t$ and $\bm{R}_t$ represent the internal inductances and resistances. Finally, $\bm{R}_{line}$ and $\bm{L}_{line}$ correspond to the inductances and resistances of the power lines connecting the DGUs. 

\begin{table}[]
    \centering
    \begin{tabular}{c|c|c|c|c|c}
       \# DGU & $R (\Omega)$ & $C_t (mF)$ & $L_t (mH)$ & $Y_{load} (m\Omega^{-1})$ &$\bar{i}_{load} (A)$  \\
        \hline
         1 &  0.2 & 2.2& 1.8 & 0.333 & 1.67  \\
         2 &  0.3 & 1.9& 2.0 & 0.370 & 2.00\\
         3 &  0.1 & 1.7& 2.2 & 0.303 & 2.33\\
         4 &  0.5 & 2.5& 3.0 & 0.303 &2.17\\
         5 &  0.4 & 2.0& 1.3 & 0.333 & 2.67\\
         6 &  0.6 & 3.0& 2.5 & 0.317 & 2.50\\
    \end{tabular}
    \caption{Electrical parameters for the DC microgrid.}
    \label{Chap2:params_grid}
\end{table}

The objective of this experiment is to design a distributed control scheme that ensures fair power sharing among the DGUs in the DC microgrid $\Sigma_{dmG}$ while regulating the weighted average voltage. This can be formulated as
% The primary goal of the control scheme is to ensure fair power sharing at steady state, which means achieving weighted power sharing based on the generation capacities of the DGUs in the network, as preset by the operator.\footnote{We assume the steady state of the DC microgrid here. Thus, implying the load current must be constant.}
a control objective to ensure that the following conditions are satisfied at the steady state:
 \begin{align}
     w_iP_i=w_jP_j, \forall \ i,j \in \mathcal{V}_s \;,
 \end{align}
where the power $P_i$ flowing out of $i^{th}$ DGU  converges to its steady state solution for all $i \in \mathcal{V}_s$, i.e., 
 \begin{align}
      \lim_{t \rightarrow \infty} P_i(t) = \bar{P}_i \;,
 \end{align}
where $\bar{P} = \bar{w}  P^\star, P^\star \in \mathbb{R}$ with $\bar{w} = [1/w_i,\dots,1/w_N]^\top, w_i > 0$ for all $i \in \mathcal{V}_s$ and any scalar $P^\star$, which is the sum of the powers drawn by each load. The value of $P^\star$ can be calculated using steady-state values of voltages and load currents, see \cite[Chapter 5]{otten2020power} for more details. For our experiment, the computed value is $P^\star = 671.57W$.
To regulate weighted average voltage, all DGUs in the network should converge to a preset reference value ${v}^\star$. In this context, the DGU with the largest capacity is assigned the highest weight, leading to the following objective:
\begin{align*}
    \lim_{t \rightarrow \infty} \sum_{i = 1}^N  \bar{w}_i {v}_i(t) = \sum_{i = 1}^N  \bar{w}_i {v}^\star \;, 
\end{align*}
where ${v}^\star = 50 V$.
Moreover, it is also desired that all the individual voltages of DGUs, i.e. $v_i(t)$ should remain within $\pm 5 \%$ of ${v}^\star$ \cite{strehle2020scalable}. Finally, since the dynamics $\Sigma_{mdG}$ are linear and incrementally stable \cite{sontag1999notions}, it is important to preserve the  $i\mathcal{L}_2$ gain of the closed-loop for good reference tracking.
Accordingly, the OCP can be formulated as follows:

\begin{align*}
     \min_{\bm{\theta}} \frac{1}{S} \sum_{k=1}^S c(\bm{x}(t), &\bm{u}(t); \bm{\theta}, \bm{x}_0^k) \  \nonumber \\
    c(\bm{x}(t),\bm{u}(t), \bm{x}_0) &= \frac{1}{2NT} \int_{0}^T \sum_{i = 1}^N \left( \Vert  \bar{w}_i {v}_i(s) - \bar{w}_i {v}^\star \Vert \right.  \\
    & \left. \qquad + \Vert {v}_i(s) {i}_{t,i}(s) - \bar{w}_i P^\star \Vert \right)ds \;,
    \end{align*}
such that $\Sigma_{dmg} \Vert_f \Sigma_{pH}$ has a finite  $i\mathcal{L}_2$ gain. For this OCP, we choose $\bar{w} = [0.3, 0.1, 0.05, 0.15, 0.22, 0.18]^\top$.  
% This means that each DGU should generate power in proportion to its maximum production capacity relative to the capacities of all DGUs in the network.
% Fundamentally, a microgrid consists of both generators and loads. To meet load demand, the generators must produce an amount of power equal to the consumption, maintaining balance within the power system. 
% A microgrid may have as many producers as loads, each with a limited generation capacity. Indeed, every power source has a maximum output capacity. This means that when a high-demand load is activated, the required power should be distributed among the available power sources. This consideration leads to the concept of power sharing. 
% The simplest form of power sharing involves each DGU supplying an equal amount of power at steady state. However, in practice, different power sources have varying capacities, making weighted power sharing more appropriate. 
% This approach assigns each producer a weight, ensuring that the power injections are balanced according to these weights. 

Note that similar optimization problems have already been addressed in several works as reviewed in  \cite{meng2017review}. Our main goal is not to outperform state-of-the-art solutions but to demonstrate power sharing and voltage balancing by solving a single OCP via the pH controller \eqref{Chap2:eq.Controller}. Moreover, our approach paves the way for the use of more complex control costs that, in addition to the above goals, can account for nonlinear phenomena, such as component aging or nonlinear loads.

In this experiment, for the pH controller \eqref{Chap2:eq.Controller},
we set $\bm{J}_c = \texttt{blkdiag}(J_i)$ where $J_i = \Gamma_i - \Gamma^\top_i$ and $\Gamma_i \in \mathbb{R}^{3 \times 3}$ are lower triangular matrices, $\bm{\Lambda} = 10 I_{18}$, $\bm{G}_c = \texttt{blkSparse}(\mathcal{P}_e)$, $\alpha = 2.0 \bar{\lambda}(\bm{G}_c \bm{G}_c^\top)$ and the Hamiltonian function 
    $H_c(\bm{\xi}) = \log \left( \cosh [ \texttt{blkdiag}(K_i) \bm{\xi} ] \right)^\top \mathds{1} + 0.1 \bm{\xi}^\top \bm{\xi}$, with $K_i \in \mathbb{R}^{3 \times 3}$. 
We choose $S=1$ in this example. However, after training, we simulated the closed-loop on different initial conditions to validate the performance. The whole training procedure takes almost $1.5$ hours on a Bizon ZX5000 G2 workstation.
See Appendix \ref{Chap2:sec:training} for more details on the training. 
% Moreover, one can also analytically compute the closed-form solution of the Jacobian of \eqref{Chap2:eq:hamiltonian_example} for each sub-controller as 
% \begin{equation*}
%     \frac{\partial H_i}{\partial \xi_i}(\xi, \theta_i) = K^\top_i \tanh (K_i \xi_i), \ \forall i \in \mathcal{V}\;.
% \end{equation*}

\begin{figure}
    \centering
    \includegraphics[width=\linewidth]{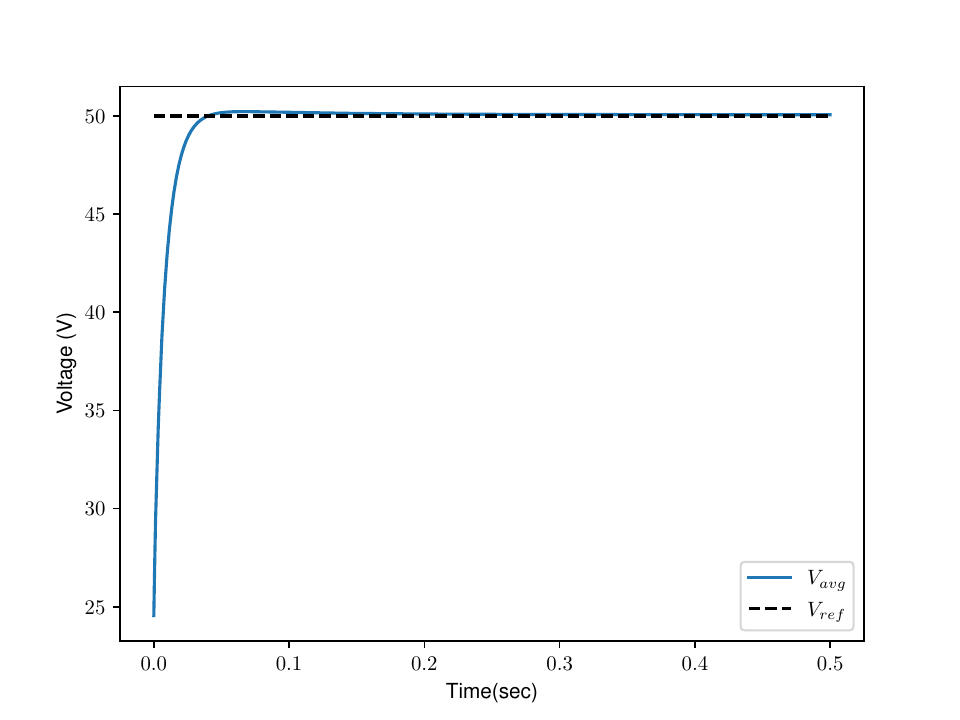}
    \caption{Averaged voltage of the whole microgrid.}
    \label{Chap2:weighted_voltages}
\end{figure}

\begin{figure}
    \centering
    \includegraphics[width=\linewidth]{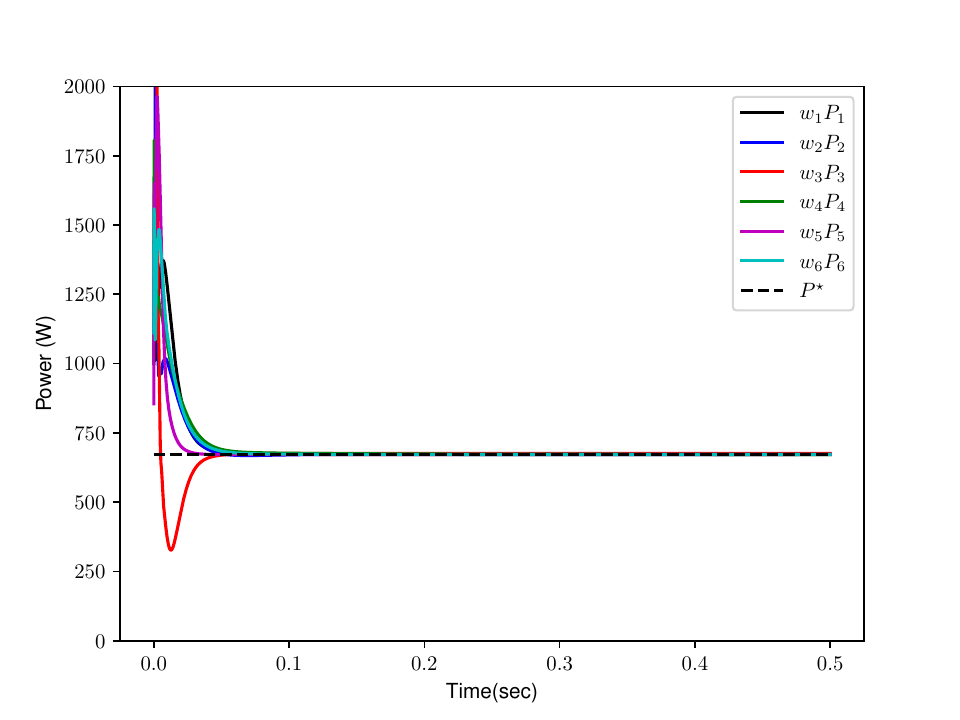}
    \caption{Weighted averaged power converging to $P^\star$.}
    \label{Chap2:weighted_powers}
\end{figure}

\begin{figure}
    \centering
    \includegraphics[width=\linewidth]{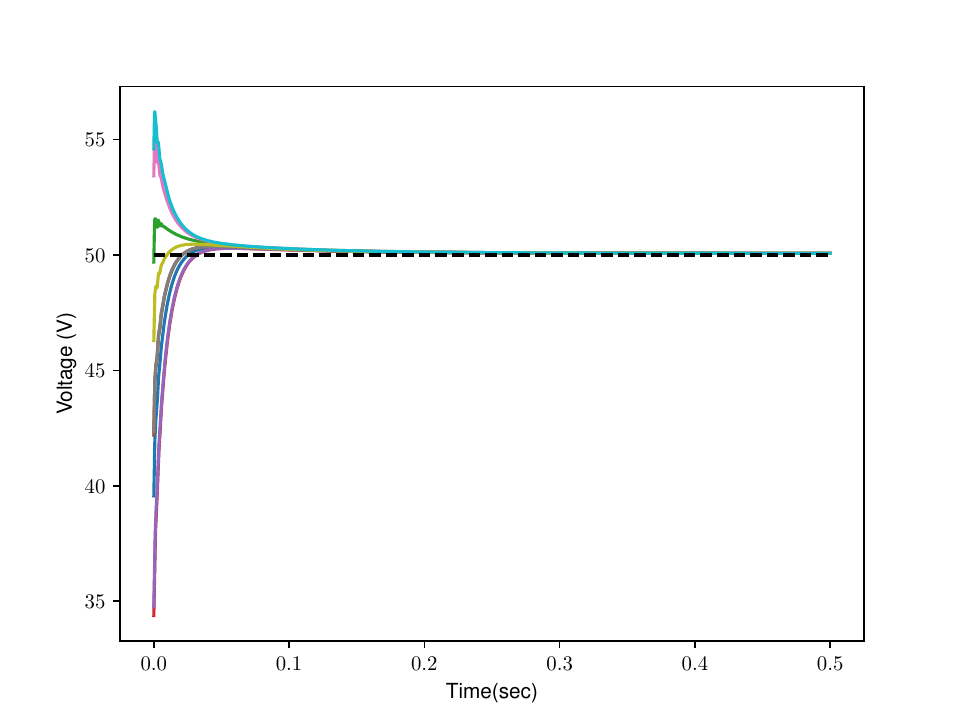}
    \caption{Averaged voltage of the whole microgrid for 10 random initial conditions.}
    \label{Chap2:weighted_voltages_randoms}
\end{figure}

\begin{figure}
    \centering
    \includegraphics[width=\linewidth]{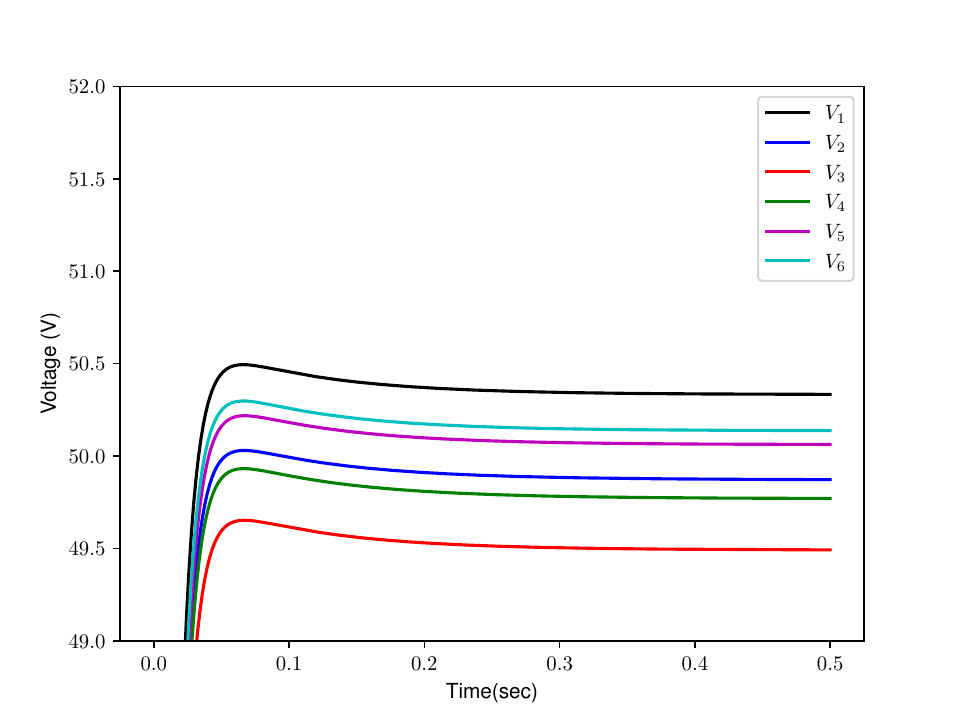}
    \caption{Voltages of individual DGUs.}
    \label{Chap2:individual_voltages}
\end{figure}

Once the training is done, we simulated the closed-loop $\Sigma_{dmG}\Vert_f\Sigma_{pH}$ and both the regulation of weighted voltage and the power-sharing are showcased in Figs. \ref{Chap2:weighted_voltages},  and \ref{Chap2:weighted_powers}, respectively.  The results indicate that the distributed pH controller has successfully achieved the desired reference voltage $v^\star$ and averaged power $P^\star$, which represents the total load requirement of the entire microgrid. Moreover, in Fig. \ref{Chap2:weighted_voltages_randoms}, we present trajectories of averaged voltages of DGUs for 10 randomly chosen initial conditions converging to $v^\star$ demonstrating good reference tracking. Finally,
Fig. \ref{Chap2:individual_voltages} displays the individual voltages at the points of connection of each DGU 
 remaining within the normal operating range of $\pm 5 \%$ around $v^\star$, hence, demonstrating the efficacy of the proposed pH-based control framework.

% \begin{figure}
%     \centering
%     \includegraphics[width = \linewidth]{EPFL_thesis_template-master/2_Control_w_Interconnection/CDC2024/Figs/consensus_w.eps}
%     \caption{Forward velocities of the swarm demonstrating a consensus.}
%     \label{fig:my_label1}
% \end{figure}
% \begin{figure}
%     \centering
%     \includegraphics[width = \linewidth]{EPFL_thesis_template-master/2_Control_w_Interconnection/CDC2024/Figs/relative_distance_w.eps}
%     \caption{Relative distance among the agents with collision avoidance.}
%     \label{fig:my_label2}
% \end{figure}
% \begin{figure}[h!]
%     \centering
%     \includegraphics[width = \linewidth]{EPFL_thesis_template-master/2_Control_w_Interconnection/CDC2024/Figs/relative_distance_wo.eps}
%     \caption{Relative distance among the agents without the collision avoidance. Note that, relative distance $||r_{ij}|| < 0.5$ means a collision.}
%     \label{fig:my_label3}
% \end{figure}

% Conclude the control (What we have proposed) and how much we 

\section{Conclusion and outlook} \label{Chap2:sec:conclusion}
The distributed control of large-scale nonlinear systems can pose several challenges, such as guaranteeing closed-loop stability. 
To tackle this issue, we have proposed an unconstrained parametrization of distributed controllers via Hamiltonian models that preserve closed-loop stability and guarantee a finite $\mathcal{L}_2$ or an $i\mathcal{L}_2$ gain, for arbitrarily large networks of nonlinear dissipative systems. We provided discretization schemes that can preserve these dissipative properties for implementation purposes. We demonstrated that near-optimal performance can be achieved by parametrizing nonlinear storage functions via some NNs for the controllers. Moreover, these NN structures can be leveraged for nonlinear system identification from data, where for example, the identified models are stable by design and have a finite $\mathcal{L}_2$ gain.

Further efforts will be devoted to implementing the controllers on real-world systems and also to generalize the framework beyond pH models.

\appendix

\subsection{Proof of Theorem \ref{Chap2:thm:finite_l2_gain}}
\label{Chap2:thm2_proof}
    The proof is done by showing that the controller \eqref{Chap2:eq.Controller}
is $\epsilon$-strictly output passive for $\alpha \geq \epsilon \bar{\lambda}(\bm{G}_c \bm{G}_c^\top)$. Recall from \cite{vanderSchaft2017}, that the controller \eqref{Chap2:eq.Controller} is $\epsilon$-output strictly passive if and only if the following conditions hold $\forall \bm{\xi} \in \Xi$
\begin{subequations}
    \begin{equation} \label{Chap2:eq.inq_1}
            \frac{\partial }{\partial \bm{\xi}}V f(\bm{\xi}) \leq  - \epsilon h^\top(\bm{\xi}) h(\bm{\xi})
    \end{equation}
    \begin{equation} \label{Chap2:eq.inq_2}
        \frac{\partial }{\partial \bm{\xi}}V g(\bm{\xi}) = h^\top(\bm{\xi}) \;,  
    \end{equation}
\end{subequations}
where $V(\bm{\xi})$ is a $\mathcal{C}^1$ storage function and satisfies $V \geq 0$. 
% Note that the control-affine system is output strictly passive if there exists $\epsilon > 0$ such that the controller \eqref{eq.Controller} is dissipative with respect to supply-rate $s(\bm{u}, \bm{y}) = \bm{u}^\top \bm{y} - \epsilon ||\bm{y}||^2$.

Moreover, $f(\bm{\xi}) = (\bm{J}_c - (\alpha \bm{I} + \bm{\Lambda})) \frac{\partial H(\bm{\xi},\bm{\vartheta})}{\partial \bm{\xi}}$, $g(\bm{\xi}) =  \bm{G}_c$, and $ h(\bm{\xi}) = \bm{G}^\top_c \frac{\partial H(\bm{\xi},\bm{\vartheta})}{\partial \bm{\xi}}$. 
First, we show that inequality \ref{Chap2:eq.inq_1} is satisfied by design.
Let us choose the candidate storage function as the Hamiltonian of controller \eqref{Chap2:eq.Controller}, i.e. $V(\bm{\xi}) = H_c(\bm{\xi},\bm{\vartheta}) \geq 0$, then we have
\begin{align*}
     &\frac{\partial^\top H(\bm{\xi}) }{\partial \bm{\xi}} f(\bm{\xi}) \leq - \epsilon h^\top(\bm{\xi}) h(\bm{\xi}), \quad \forall \bm{\xi} \in \Xi  \\
      &\frac{\partial^\top H(\bm{\xi}) }{\partial \bm{\xi}} (\bm{J} - (\alpha \bm{I} + \bm{\Lambda})) \frac{\partial H(\bm{\xi})}{\partial \bm{\xi}} \leq - \epsilon  \frac{\partial^\top H(\bm{\xi}) }{\partial \bm{\xi}} \bm{G}_c \bm{G}^\top_c \frac{\partial H(\bm{\xi})}{\partial \bm{\xi}} \\
      &\frac{\partial^\top H(\bm{\xi}) }{\partial \bm{\xi}} \bigg( -\alpha \bm{I} - \bm{\Lambda} + \epsilon  \bm{G}_c \bm{G}^\top_c   \bigg) \frac{\partial H(\bm{\xi})}{\partial \bm{\xi}} \leq 0 \; .
\end{align*}
Therefore, choosing  $\alpha \geq \epsilon \bar{\lambda}(\bm{G}_c \bm{G}_c^\top)$ verifies the last inequality.
Note that the second equality \eqref{Chap2:eq.inq_2} is verified by construction due to the choice of storage function and the structure of the pH controller \eqref{Chap2:eq.Controller}. 
Finally, by employing \cite[Theorem 2.2.13]{vanderSchaft2017} we conclude that if the controller \eqref{Chap2:eq.Controller} is $\epsilon$-output strictly passive, then it has a finite $\mathcal{L}_2$-gain $ \leq 1/\epsilon$ for all trainable parameters $\bm{\theta}$. \hfill $\square$ 

\subsection{Proof of Theorem \ref{thm:incremental_l2_gain}}
\label{Chap2:thm2:inc_proof}
    First, let us show that differential controller \eqref{Chap2:diff_ph} is  $\epsilon_\delta$-output strictly incrementally passive with a suitable storage function. 
To this aim, recall from \cite{vanderSchaft2017}, that the controller \eqref{Chap2:eq.Controller} is strictly output  passive if and only if the following conditions hold $\forall \bm{\xi}$
\begin{subequations}
    \begin{equation} \label{eq.inequality_1}
          \frac{\partial }{\partial \bm{\xi}}V_\delta f(\bm{\xi}, \delta \bm{\xi}) \leq  - \epsilon_\delta h^\top(\bm{\xi},\delta \bm{\xi}) h(\bm{\xi},\delta \bm{\xi}), 
    \end{equation}
    \begin{equation} \label{eq.inequality_2}
    \frac{\partial }{\partial \bm{\xi}}V_\delta g(\bm{\xi},\delta \bm{\xi}) = h^\top(\bm{\xi},\delta \bm{\xi}) \;, 
    \end{equation}
\end{subequations}
where $V_\delta(\bm{\xi},\delta\bm{\xi})$ is a storage function for variational dynamics \eqref{Chap2:diff_ph} and $\epsilon_\delta > 0$ is a positive constant. Moreover, 
\begin{align*}
    f(\bm{\xi},\delta \bm{\xi}) &= (\bm{J} - (\alpha \bm{I} + \bm{\Lambda})) \frac{\partial^2 H_{c}(\bm{\xi},\bm{\vartheta}) }{\partial \bm{\xi}^2} \delta \bm{\xi}, \quad g(\bm{\xi},\delta \bm{\xi}) =  \bm{G}_c \\ 
    h(\bm{\xi},\delta \bm{\xi}) &= \bm{G}^\top_c \frac{\partial^2 H_{c}(\bm{\xi},\bm{\vartheta}) }{\partial \bm{\xi}^2} \delta \bm{\xi} \;,
\end{align*}
 respectively. 
Consider the storage function $\displaystyle{V_\delta(\bm{\xi}, \delta \bm{\xi})= \frac{1}{2} \delta \bm{\xi}^\top \frac{\partial^2 H(\bm{\xi},\bm{\vartheta}) }{\partial \bm{\xi}^2} \delta \bm{\xi} }$, that verifies $V_\delta(\cdot,0) = 0$, and $V_\delta(\cdot,\cdot) > 0$. The latter inequality can be verified by the condition \ref{Chap2:hessian_cond}. Then, for the inequality \eqref{eq.inequality_1}, we obtain
    \begin{align*}
     &\delta \bm{\xi}^\top  \frac{\partial^2 H(\bm{\xi})^\top }{\partial \bm{\xi}^2} (\bm{J} - (\alpha \bm{I} + \bm{\Lambda})) \frac{\partial^2 H(\bm{\xi}) }{\partial \bm{\xi}^2} \delta \bm{\xi} \leq \\
      & \hspace{2cm}-\epsilon_\delta \delta \bm{\xi}^\top \frac{\partial^2 H(\bm{\xi})^\top }{\partial \bm{\xi}^2} \bm{G}_c \bm{G}^\top_c \frac{\partial^2 H(\bm{\xi}) }{\partial \bm{\xi}^2} \delta \bm{\xi} \;.
\end{align*}
Thus, by choosing $\alpha \geq  \epsilon_\delta \bar{\lambda}(\bm{G}_c\bm{G}^\top_c)$, the differential controller is $\epsilon_\delta$-output strictly incrementally passive and the finite $i\mathcal{L}_2$ gain is less than $1/ \epsilon_\delta$ \cite[Proposition 2.2.21]{vanderSchaft2017}. 
Finally, the incremental dissipativity of the controller \eqref{Chap2:eq.Controller} with $\alpha \geq \epsilon_\delta \bar{\lambda}(\bm{G}_c\bm{G}^\top_c)$ follows from \cite[Theorem 6]{verhoek2023convex}. \hfill $\square$

\subsection{Proof of Theorem \ref{Chap2:thm:discrete_L2_gain}}
\label{Chap2:proof_of_thm_5}

    Our goal is to show that
    \begin{align}
          \frac{H_c(\bm{\xi}_{k+1}) - H_c(\bm{\xi}_{k})}{h_\Delta}   \leq s(\bm{u}_k,\bm{y}_k)    
    \end{align}
    where $s(\bm{u}_k,\bm{y}_k)$ is a supply-rate, and $h_\Delta>0$ is an arbitrary step-size. 
    Let us consider the following supply rate \cite{vanderSchaft_overview}
    \begin{align*}
        s(\bm{u}_k,\bm{y}_k) =  \bm{u}^\top_k \bm{y}_k - \epsilon ||\bm{u}_k||_2^2 \;.
    \end{align*}
    Then, by using the definition of discrete gradient (Definition \ref{Chap2:def:discrete_gradients}), we obtain
    \begin{align*}
      {H_c(\bm{\xi}_{k+1}) - H_c(\bm{\xi}_{k})}    &=   \bar{\nabla} H_c(\bm{\xi}_{k},\bm{\xi}_{k+1})^\top (\bm{\xi}_{k+1}- \bm{\xi}_{k})  \\ 
        &\hspace{-15mm}= \bar{\nabla} H_c^\top \left( h_\Delta(\bm{J}_c - (\alpha \bm{I} + \bm{\Lambda})) \bar{\nabla}H_c + h_\Delta\bm{G}_c \bm{y}_k  \right) \\ 
        &\hspace{-15mm}= \bar{\nabla} H_c^\top \left(- h_\Delta (\alpha \bm{I} + \bm{\Lambda}) \bar{\nabla}H_c + h_\Delta\bm{G}_c \bm{y}_k  \right)  \\
        &\hspace{-15mm}\leq h_\Delta \bm{u}^\top_k \bm{y}_k - h_\Delta \epsilon ||\bm{u}_k||_2^2 \;.
    \end{align*}
    By setting $\bm{u}_k =  \bm{G}^\top_c \bar{\nabla}H_c$, we can cancel $h_\Delta  \bar{\nabla}H_c^\top \bm{G}_c \bm{y}_k$ on both sides and then, after dividing the inequality by $h_\Delta > 0$, we get  
    \begin{align*}
         \bar{\nabla} H_c^\top \left(- (\alpha \bm{I} + \bm{\Lambda}) \bar{\nabla}H_c + \bm{G}_c \bm{y}_k  \right)  
        \leq  \bm{u}^\top_k \bm{y}_k -  \epsilon ||\bm{u}_k||_2^2 \;.
    \end{align*}
    % for the above inequality, we have shown that Theorem \ref{Chap2:thm:finite_l2_gain}, that it is satisfied for $\alpha \geq \epsilon \bar{\lambda}(\bm{G}_c \bm{G}_c^\top)$.
    Therefore, by choosing $\alpha \geq \epsilon \bar{\lambda}(\bm{G}_c \bm{G}_c^\top)$ and discretizing through a discrete gradient method, one can preserve the finite $\mathcal{L}_2$ gain.
\hfill $\square$

\subsection{Dissipation-preserving discretization of pH controllers} \label{subsec:disspation_preserving} 
In this investigation, we demonstrate discretization using discrete-gradient methods, such as the Itoh-Abe method \cite{ehrhardt2018geometric}, preserves the $\mathcal{L}_2$ gain, while some common discretization schemes such as Forward Euler do not. 

\begin{figure}
    \centering
    \includegraphics[width=\linewidth]{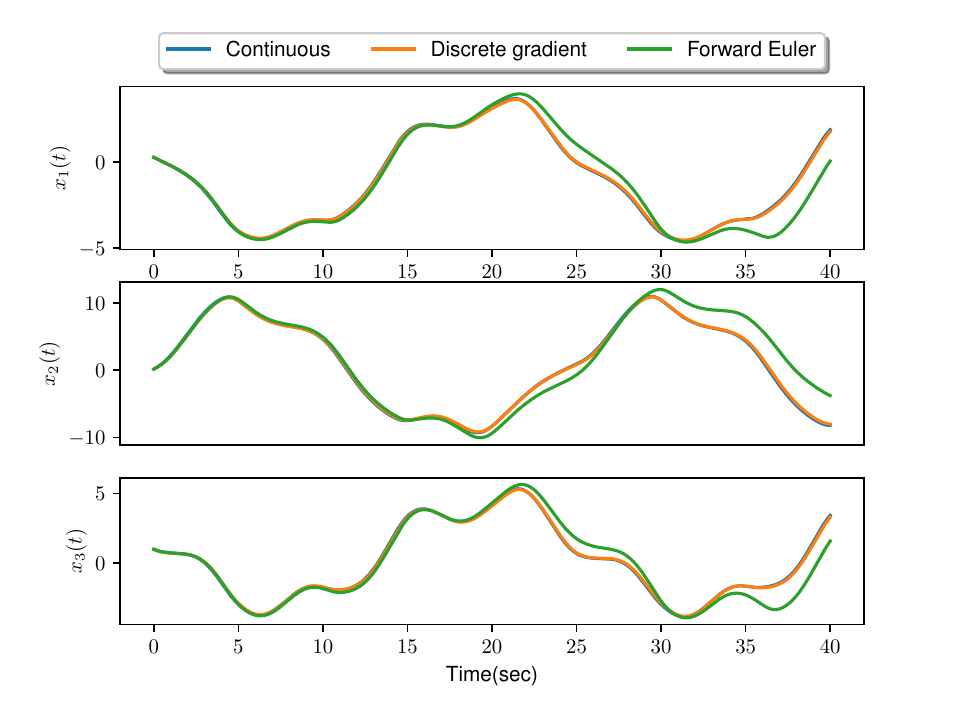}
    \caption{Evolution of trajectories for a random initial condition.}
    \label{Chap2:states_comp}
\end{figure}

\begin{figure}
    \centering
    \includegraphics[width=\linewidth]{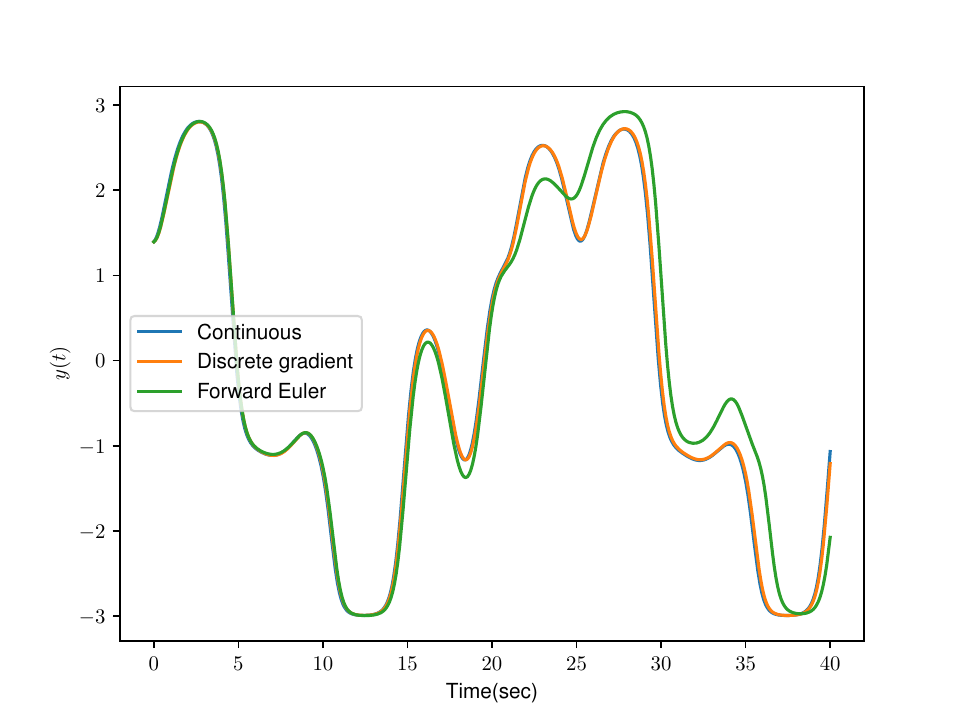}
    \caption{The comparison of the outputs of the pH controllers subject to a sinusoidal input $u(t) = \sin(t)$. }
    \label{Chap2:fig:output_comp}
\end{figure}

Consider the continuous-time pH controller $\Sigma_{pH}$ \eqref{Chap2:eq.Controller} with parameters
\begin{align*}
    J_c = \begin{bmatrix}
        0 & -1 & 0 \\ 1 & 0  & 1 \\ 0 & -1 & 0
    \end{bmatrix}, \ G_c = \begin{bmatrix}
        0.5 \\ 1.0 \\ 1.0
    \end{bmatrix} \;.
\end{align*}
Moreover, set $\alpha = 0.001 \bar{\lambda}(G_c G_c^\top)$ (See Theorem \ref{Chap2:thm:finite_l2_gain}), and $H_c(\xi)= \log (\cosh(K \xi))^\top \mathds{1}$, where $K \in \mathbb{R}^{3 \times 3}$ is chosen randomly. We discretize $\Sigma_{pH}$ with the above parameters using the Itoh-Abe method \cite{ehrhardt2018geometric} to obtain \eqref{Chap2:eq:discrete_cont} and compare the results with the simple Forward Euler (FE). In both schemes, we chose a step size of $h = 0.1$ seconds. The states and outputs of all controllers, i.e., continuous, discrete gradients, and FE, are plotted in Fig. \ref{Chap2:states_comp} and Fig. \ref{Chap2:fig:output_comp}, respectively. We observe that the trajectories of the continuous-time and discrete-time controllers using the Itoh-Abe method are nearly overlapping, whereas the trajectories of the controller based on FE method diverge. This comparison demonstrates the importance of using discrete-gradient methods for the discretization of dissipative controllers.

\subsection{Training of port-Hamiltonian controllers}
\label{Chap2:sec:training}
One can solve the nonlinear OCP \eqref{Chap2:opt_prob}-\eqref{Chap2:const:l2} by casting it as a NN training problem as follows \cite{NeuralODEs}, 

\begin{align}
  & \min_{ \{\bm{\theta}_k\}_{k = 0}^L}\quad \frac{1}{S} \sum_{i=1}^S c(\bm{x},\bm{u};\bm{\theta}_t,\bm{z}_0^k)  \nonumber \\
 & \begin{aligned}
  \mathrm{s.t.} \quad 
  &\bm{z}_0^i,\dots,\bm{z}_L^i = \operatorname{ODESolve}\{\bm{f}(\bm{z}_t,\bm{\theta}_t,t), \bm{z}_0^i, (t_0,\dots,t_L)\}\;, \nonumber 
  \end{aligned} 
\end{align}
where $\bm{\theta}_k$ are the samples of $\bm{\theta}_t$ at time steps $(t_0, \ldots, t_L)$, $\bm{z}_i = [ \bm{x}_i, \bm{\xi}_i ]$ are stacked states of the distributed system and the controller. Moreover, ODESolve is any numerical solver that returns the sampled closed-loop trajectory $\bm{z}_0^i,\dots,\bm{z}_L^i$  of the closed-loop $\bm{f}(\cdot):=\Sigma_s \Vert_f \Sigma_{pH}$ starting from an initial condition $\bm{z}_0^i$, and  $L$ is the number of sampled states. Several ODE solvers can be used to simulate the dynamics, such as Foward Euler, or Dopri5 \cite{poli2020torchdyn}. Once the closed-loop trajectories are obtained, one can evaluate the loss function ${c}$, and leverage BPTT to update the parameters via algorithms like Stochastic Gradient Descent (SGD) or \emph{Adam} \cite{kingma2015adam}.   
For details on training procedures, we refer the interested reader to \cite{NeuralODEs}.

\begin{remark}[Computation of $\alpha$ through BPTT]
\label{Chap2:remark_compute_alpha}
 During BPTT, the following procedure can be adopted to ensure that 
$\Sigma_{pH}$ remains dissipative throughout the optimization process:
Perform forward propagation through the closed-loop system 
$\bm{f}(\cdot)$ using the current value of 
$\bm{\vartheta}$. Compute 
$\alpha$ according to either Theorem \ref{Chap2:thm:finite_l2_gain} or \ref{thm:incremental_l2_gain} for a batch of given initial conditions, and then calculate the training loss.
  ii) update the parameters $\bm{\vartheta}$ via BPTT, and iii)  Recompute
$\alpha$ using the updated parameters.
\end{remark}

% Note that the closed-loop dynamics 
% $\Sigma_{s} \Vert_f \Sigma_{pH}$ form a continuous-time Hamiltonian system with trainable parameters 
% $\{K_i(t) , b_i(t)\}$ There are several methods to train these weights; one popular approach is to discretize the closed-loop dynamics using an appropriate scheme, such as the FE method. The resulting discrete-time system can then be interpreted as a deep neural network, allowing the use of standard algorithms, such as backpropagation through time, to optimize the parameters. 
% For instance, applying FE on $\dot{x}(t) = {f}({x}(t),{\bm{\theta}(t)})$, one obtains the following discrete-time system  
% \begin{equation*}
%     x_{k+1} = x_k + h f(x_k,\bm{\theta}_k), \ \text{for} \ k = 0,1,\cdots, N, 
% \end{equation*}
% which can be seen as the forward equation for well-known Residual networks (ResNet) \cite{he2016deep} with $N$ number of layers. Therefore, standard packages such as PyTorch \cite{pytorch} can be used. 

\subsubsection{Implementation details for Experiment \ref{Chap2:example1}}

We simulate the closed-loop using FE with $h = 0.0625$ for a finite horizon of $T = 5$ secs. We train our controller for $1000$ epochs.  Particularly,  we chose {Adam} \cite{kingma2015adam} as the optimizer with a learning rate of 0.01 for training. 

\subsubsection{Implementation details for Experiment \ref{Chap2:sec:example2}} 

The closed-loop is simulated using FE with a step-size of $h = 2.5e-5$ for $T = 0.5$ secs. We train our controller with  $2500$ epochs using the {Adam} optimizer \cite{kingma2015adam} with a learning rate of $5 \times 10^{-3}$. 

We highlight that discretization during training might lead to sub-optimality. However, it does not compromise the closed-loop stability guarantees from Theorem \ref{Chap2:thm:small_gain}, and \ref{Chap2:thm2:inc_stab}. 
This is because the $\mathcal{L}_2$ ($i\mathcal{L}_2$) gain of  continuous-time system $\Sigma_{pH}$ holds regardless of the weights, as long as $\alpha$ is chosen as in Theorem \ref{Chap2:thm:finite_l2_gain}, or \ref{thm:incremental_l2_gain}.

% We highlight that discretization during training might lead to sub-optimality. However, it does not compromise the closed-loop stability guarantees from Theorem \ref{Chap2:thm:small_gain}. 
% This is because the $\mathcal{L}_2$ or incremental $\mathcal{L}_2$ gain of the continuous-time system $\Sigma_{pH}$ holds regardless of the weights, as long as $\alpha$ is chosen as in Theorem \ref{Chap2:thm:finite_l2_gain}, or Theorem \ref{thm:incremental_l2_gain}. 

\bibliography{bibliography}
\bibliographystyle{IEEEtran}

\begin{IEEEbiography}[{\includegraphics[width=1in,height=1.25in,clip,keepaspectratio]{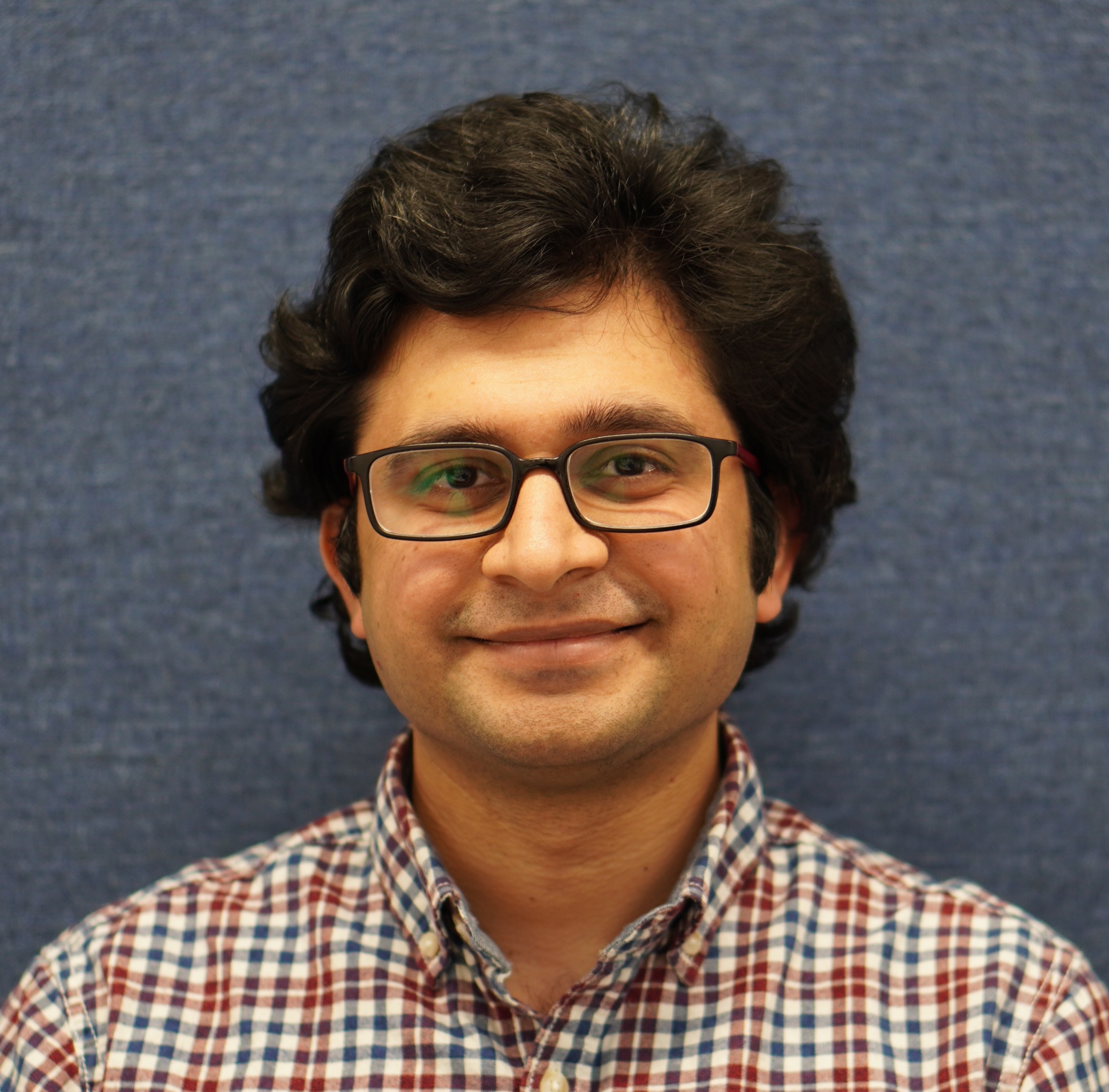}}]{Muhammad Zakwan}
 is a doctoral assistant in the Dependable Control and Decision group (DECODE) at École Polytechnique Fédérale de Lausanne (EPFL). He is also a member of the National Centre of Competence in Research (NCCR) Automation. He holds an Electrical and Electronics  Engineering degree from Bilkent University (Turkey) and a Bachelor of Science in Electrical Engineering from the Pakistan Institute of Engineering and Applied Sciences (Pakistan). His research interests include neural networks, nonlinear control, contraction theory,  port-Hamiltonian systems, and machine learning.
\end{IEEEbiography}
\begin{IEEEbiography}[{\includegraphics[width=1in,height=1.25in,clip,keepaspectratio]{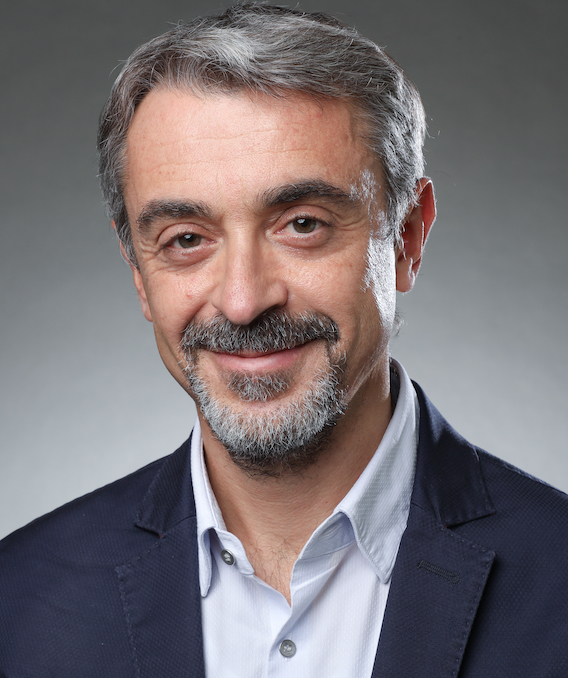}}]{Giancarlo Ferrari Trecate}
%Giancarlo Ferrari Trecate 
(SM'12) received a Ph.D. in Electronic and Computer Engineering from the Universita' Degli Studi di Pavia in 1999. Since September 2016, he has been a Professor at EPFL, Lausanne, Switzerland. In the spring of 1998, he was a Visiting Researcher at the Neural Computing Research Group, University of Birmingham, UK. In the fall of 1998, he joined the Automatic Control Laboratory, ETH, Zurich, Switzerland, as a Postdoctoral Fellow. He was appointed Oberassistent at ETH in 2000. In 2002, he joined INRIA, Rocquencourt, France, as a Research Fellow. From March to October 2005, he was a researcher at the Politecnico di Milano, Italy. From 2005 to August 2016, he was Associate Professor at the Dipartimento di Ingegneria Industriale e dell'Informazione of the Universita' degli Studi di Pavia.
His research interests include scalable control, machine learning, microgrids, networked control systems, and hybrid systems.
Giancarlo Ferrari Trecate is the founder and current chair of the Swiss chapter of the IEEE Control Systems Society. He is Senior Editor of the IEEE Transactions on Control Systems Technology and has served on the editorial boards of Automatica and Nonlinear Analysis: Hybrid Systems.
\end{IEEEbiography}

\end{document}